\documentclass[twocolumn]{aastex63}

\usepackage{amsmath}
\usepackage{xcolor}
\shorttitle{EXOMOON POLLUTION}
\shortauthors{}

\graphicspath{{./Figures/}}

\begin{document}

\title{EXOMOONS AS SOURCES OF WHITE DWARF POLLUTION}

\author{Isabella L. Trierweiler}

\affiliation{Department of Earth, Planetary, and Space Sciences, UCLA,
Department of Physics and Astronomy, University of California, Los Angeles, CA 90095-1562}

\correspondingauthor{Isabella L. Trierweiler}
\email{itrierweiler@g.ucla.edu}

\author{Alexandra E. Doyle}
\affiliation{Department of Earth, Planetary, and Space Sciences\\
University of California, Los Angeles\\
Los Angeles, CA 90095, USA}

\author{Carl Melis}
\affiliation{Center for Astrophysics and Space Sciences, University of California, San Diego, CA 92093-0424, USA}
\email{cmelis@ucsd.edu}

\author{Kevin J. Walsh}
\affiliation{Southwest Research Institute
1050 Walnut St. Suite 400
Boulder, CO 80302}

\author{Edward D. Young}
\email{eyoung@epss.ucla.edu}
\affiliation{Department of Earth, Planetary, and Space Sciences\\
University of California, Los Angeles\\
Los Angeles, CA 90095, USA}

\begin{abstract}
 Polluted white dwarfs offer a unique way to study the bulk compositions of exoplanetary material, but it is not always clear if this material originates from comets, asteroids, moons, or planets. We combine N-body simulations with an analytical model to assess the prevalence of extrasolar moons as white dwarf (WD) polluters. Using a sample of observed polluted white dwarfs we find that the extrapolated parent body masses of the polluters are often more consistent with those of many solar system moons, rather than solar-like asteroids. We provide a framework for estimating the fraction of white dwarfs currently undergoing observable moon accretion based on results from simulated white dwarf planetary and moon systems. Focusing on a three-planet white dwarf system of Super-Earth to Neptune-mass bodies, we find that we could expect about one percent of such systems to be currently undergoing moon accretions as opposed to asteroid accretion. 
\end{abstract}



\section{Introduction}\label{section:intro}
 White dwarfs (WDs) are the end-states of medium mass stars ($M_* \lesssim 8 M_\odot$), and their high gravity leads to rapid sinking of any elements heavier than helium. This sinking occurs on the timescales of days to millions of years \citep{Koester2009, Blouin2018}, and generally leaves the spectra of white dwarfs devoid of metal features. Nevertheless, over a thousand white dwarfs have been observed to be `polluted' with heavy elements \citep{Coutu2019}, with estimates that up to half of all white dwarfs are polluted \citep{Koester2014}. The source of the pollution is believed to be the remains of rocky parent bodies, which survived the post-main sequence evolution of the host star. Surviving planets can scatter these bodies onto highly eccentric orbits, such that the objects approach the white dwarf and are tidally disrupted and subsequently accreted by the star \citep[e.g.,][]{Debes2002, Jura2003}. 

The majority of WD polluters appear rocky \citep{Doyle2020, Swan2019} and a few appear icy \citep{Hoskin2020, Farihi2013}. Many accreted bodies  are chondritic in composition, and based on this and their apparent masses, it has been often assumed that these parent bodies were small rocky bodies analogous to the asteroids, or Kuiper Belt objects \citep{Xu2017}, in the solar system. Here we attempt to quantify the fraction of polluting bodies that are exomoons rather than asteroids based on the parent body masses required for observed WDs and numerical simulations of the frequency of each population's accretion. We are motivated by the recent discovery of beryllium in a polluted white dwarf \citep{klein2021} and the interpretation that the observed large excess in Be relative to other rock-forming elements is a tell-tale indicator that the parent body accreted by the WD was an icy moon. In this interpretation, the excess Be is the result of irradiation of the icy moon in the radiation belt of its host giant planet  \citep{Doyle2021}.  \cite{Payne2016} and \cite{Payne2017} showed that moons can be liberated by close encounters between planets and that liberated moons could be accreted by a white dwarf. Here we examine this proposal in greater detail using a statistical analysis and N-body simulations of the liberation and accretion of moons.

The probability of observing a moon versus an asteroid depends on the frequency and duration of the respective accretion events relative to the detectable amount of pollution on a typical WD. Based on occurence rates of dust around A-type stars, debris belts are expected to be ubiquitous amongst polluted white dwarf progenitors \citep{Melis2016}. Accordingly, assuming that any planetary systems with moons available for accretion would also have a debris belt available, the probability of observing moon versus asteroid accretion in a given WD system is
\begin{equation}
    \frac{P_{\mbox{moon accretion}}}{P_{\mbox{asteroid accretion}}} = \frac{T_{\rm moons}}{T_{\rm asteroids}},
\end{equation}\label{Eqn:Pmoons}

\noindent where $T$ is the fraction of time that the associated population provides observable pollution. Because we are interested in cumulative times, $T$ is dependent on the accretion rate of each population ($T=T(\mbox{accretion rate)}$).

Because asteroids are expected to accrete much more frequently  than moons due to their sheer number, ${P_{\mbox{moon accretion}}}/{P_{\mbox{asteroid accretion}}}$ will depend on whether the accumulation of successive asteroid accretions is sufficient to sustain high masses of WD pollution, compared to the single accretion events of relatively larger-mass moons. 

In this paper we estimate the parameters of Equation  \ref{Eqn:Pmoons} for a system of three super-Earth to Neptune mass planets. A number of planetary architectures are capable of becoming unstable and aiding to feed meterial to the white dwarf \citep[e.g.,][]{Maldonado2022, Stephan2017, Veras2015}, however, we focus on this particular system because existing studies of asteroid (debris) belts in this architecture provide a baseline against which to  compare moon accretions. 

Our paper is organized around Equation \ref{Eqn:Pmoons}. In \textsection \ref{Jura Model} we introduce an analytical model for masses of polluting elements in the WD convection zone, which we apply to all accretion events throughout this work. We show that this model implies that observed polluted WDs require masses much larger than one would expect for typical asteroid belt objects, emphasizing the need to understand moon accretions. In \textsection \ref{section:observability} we use observations of polluted WDs to put limits on the levels of pollution required for the accreter to be detectable. \textsection \ref{Section:Asteroids} finds $T_{\rm asteroids}$, the cumulative timescale of detectability of asteroid pollution, using extrapolated asteroid accretion frequencies from previous studies. We present the results from our own N-body simulations for moon accretions in \textsection \ref{Section:nbody}, and find $T_{\rm moons}$. Finally, we summarize all quantities and discuss the implications of Equation \ref{Eqn:Pmoons} in \textsection \ref{Section:Summary}.

\section{Analytical Model for Convection Zone Masses}\label{Jura Model}
To determine the duration and pollution levels associated with an accretion event, we make use of the model by \cite{Jura2009} for the buildup of accreted material in a white dwarf atmosphere. Throughout this work we will refer to the model as J09. The model describes the time-dependent mass of the polluter currently observable in the convection zone of a white dwarf as a function of polluting parent body mass, element settling times through the WD atmosphere, and the duration of the accretion event. The model assumes the accretion disk, and therefore accretion rate, decays exponentially as one might expect from dissipative forces that depend on mass. Under this assumption, the mass of element $Z$ that is observed to be in the convection zone of the white dwarf at the time of observation $t$ after the start of the accretion event, $M_{CV}(Z,t)$, is:

\begin{equation}\label{Eqn:MCV}
    M_{\rm CV}(Z,t) = \frac{M_{\rm PB}(Z) \tau_{\rm set}(Z)}{\tau_{\rm disk} - \tau_{\rm set}(Z)} \left( e^{-t/\tau_{\rm disk}} - e^{-t/\tau_{\rm set}(Z)}\right),
\end{equation}

\noindent where $M_{\rm PB}(Z)$ is the mass of element $Z$ in the parent body, $\tau_{set}(Z)$ is the  e-folding settling time of element $Z$, and $\tau_{\rm disk}$ is the characteristic lifetime of the accretion disk. 

\begin{figure}
\centering
    \includegraphics[width=0.5\textwidth]{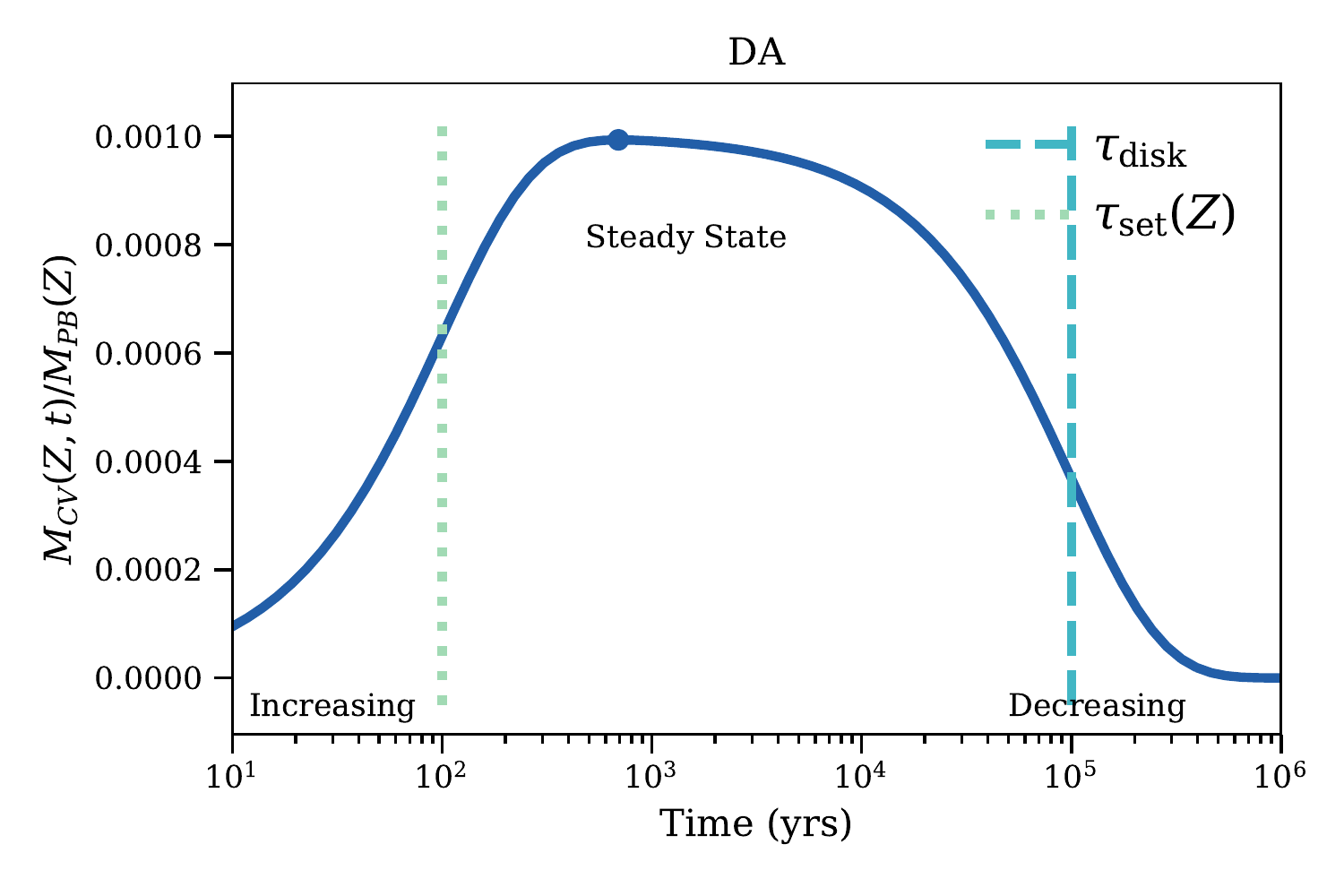}    
    \includegraphics[width=0.5\textwidth]{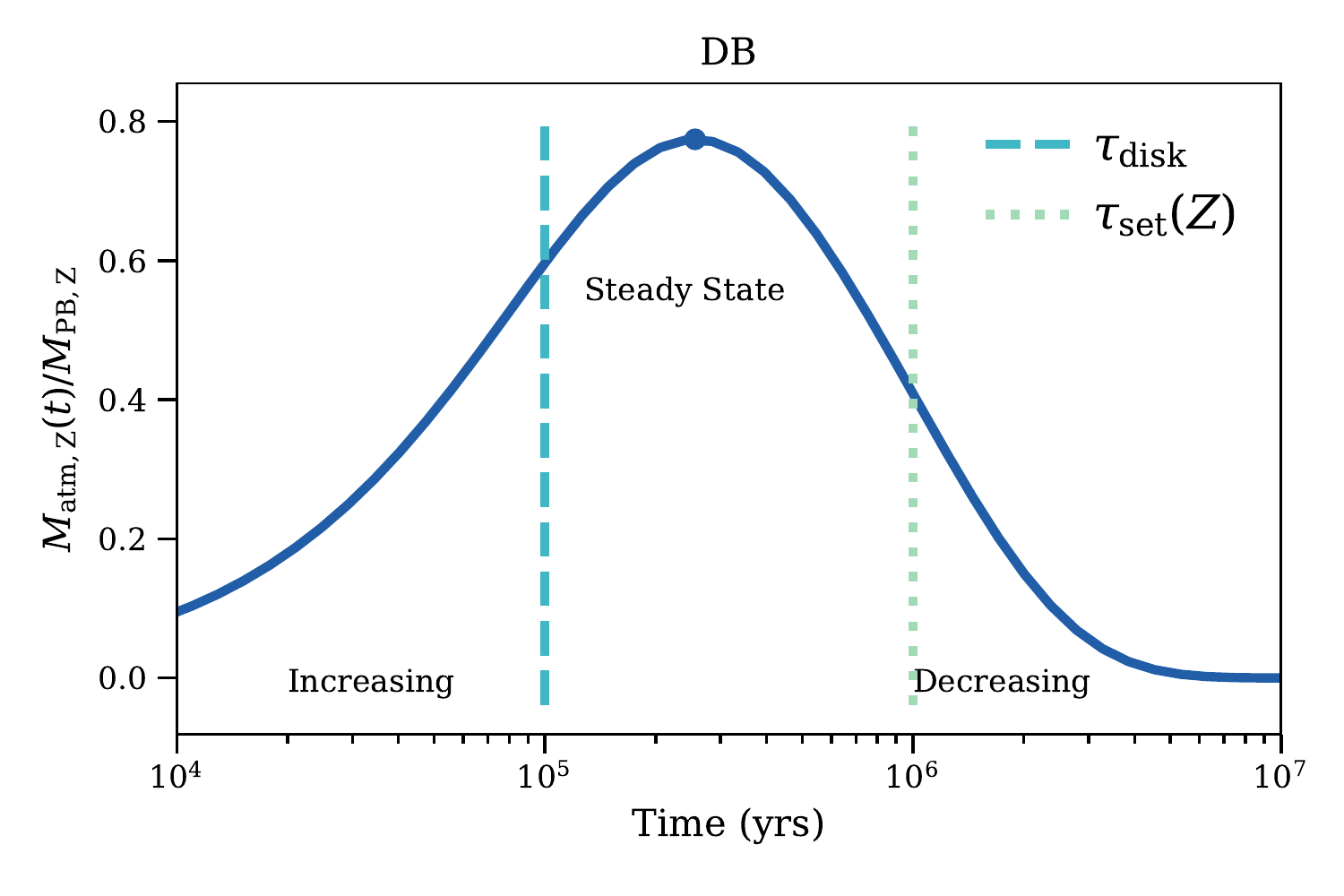} 
   \caption{Examples of the fraction of parent body masses of a particular element that are predicted to be in the convection zone of a WD as a function of time, according to the J09 model (Equation \ref{Eqn:MCV}). We assume an accretion disk lifetime of $10^5$ yr and settling times of $10^2 \rm yr$ for the DA (top) and $10^6 \rm yr$ for the DB (bottom). Note the three phases of accretion: increasing, steady state, and decreasing. While analogous phases occur both for the DA and DB, the timescales defining the boundaries of each phase are swapped due to the longer DB settling times. }\label{Fig:MCV_example}
\end{figure}

The observable pollution mass $M_{\rm CV}(Z,t)$ in J09 depends on the settling timescale, which in turn depends on the properties of the host star. Therefore, variations in white dwarf temperature and composition have significant impacts on the maximum accumulations of pollution in the WD atmosphere and the timescales during which pollution levels are sufficiently high to be observable. In Figure \ref{Fig:MCV_example} we show the pollution masses for a representative element and parent body in a DA (top) and a DB (bottom) white dwarf in order to illustrate these differences. Defined spectroscopically, DAs have atmospheres dominated by hydrogen, while those of DB white dwarfs are helium-dominated. While white dwarf classifications extend well beyond these two categories, for simplicity we will restrict our discussion to these two broad categories, using the terms DA and DB to mean hydrogen and helium-dominated in what follows. The primary difference between the two types of WDs in the present context is that DAs generally have settling timescales for the heavy elements of days to thousands of years while the DBs have settling timescales of $10^5$ to $10^6$ yr for these elements. Additionally, because hot DAs will have minimal, if any, convection zones, we consider $M_{\rm CV}$ to more generally represent the mass of observable pollution in the WD atmosphere for these cases. 

The maximum heavy element mass in the convection zone for each WD type occurs where settling times are approached. In other words, the maximum is the amount of accreted material that can build up in the atmosphere before sinking exerts an influence on the abundances. In the example to follow, we assume a settling time of $10^2$ yr for the DA and $10^6$ yr for the DB. Both cases are assigned an accretion disk e-folding lifetime of $10^5$ yr. 

In Figure \ref{Fig:MCV_example}, we show the fraction of the parent body mass that is currently observable in the convection zone as a function of time since the start of accretion, for an exemplary heavy element. Both cases illustrate the three phases of pollution: while accretion is ongoing and before settling begins, the mass of the element builds up in the atmosphere (increasing phase). When the settling and accretion rates equalize (the blue point in Figure \ref{Fig:MCV_example}), the mass stays relatively constant (steady state). Once  settling dominates as accretion wanes, the mass of the pollution decreases rapidly (decreasing phase). Note that, in this model, the majority of a single accretion event is in a decreasing phase.  

Because settling in the DBs begins after the majority of the parent body has been accreted onto the white dwarf, DBs can exhibit much larger fractions of the polluting metals than DAs, for the same parent body mass. The DA steady state phase occurs earlier and therefore at relatively higher settling and accretion rates than the DB phase, and the two DA rates conspire to cause very small fractions of the parent body to be observable at any given time. 

\begin{figure}
\centering
    \includegraphics[width=0.5\textwidth]{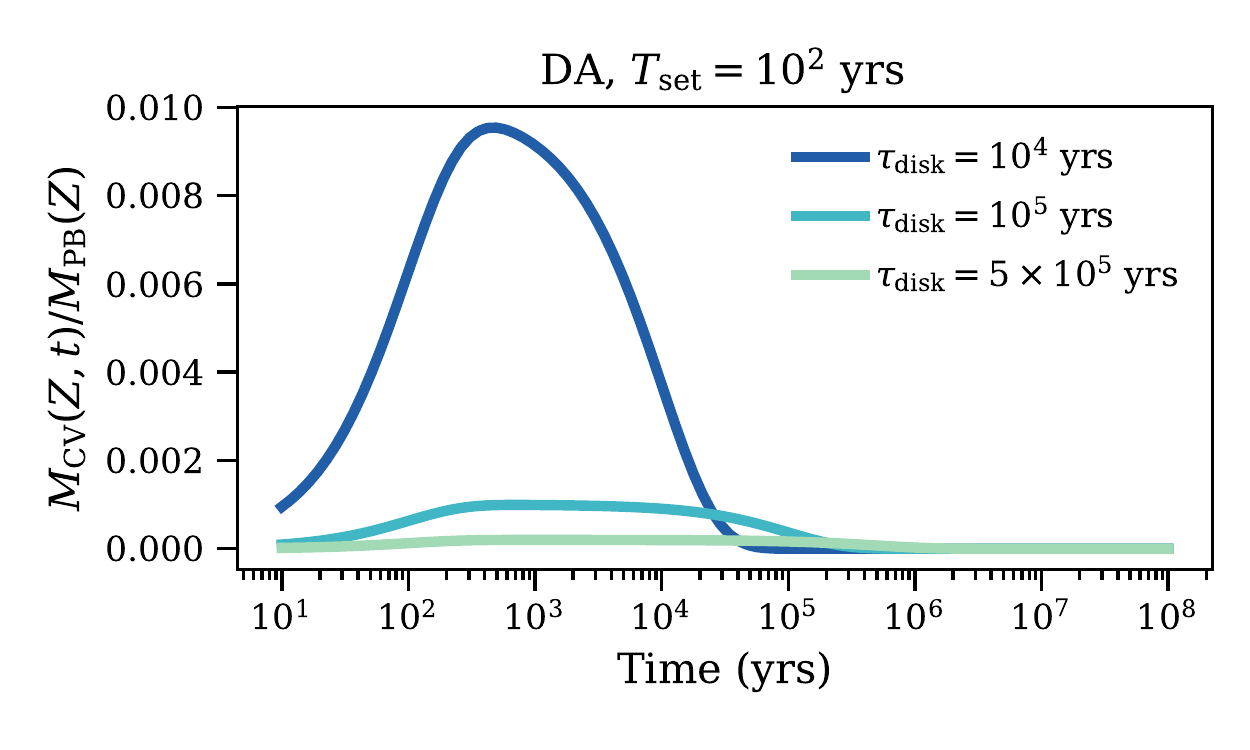}
    \includegraphics[width=0.5\textwidth]{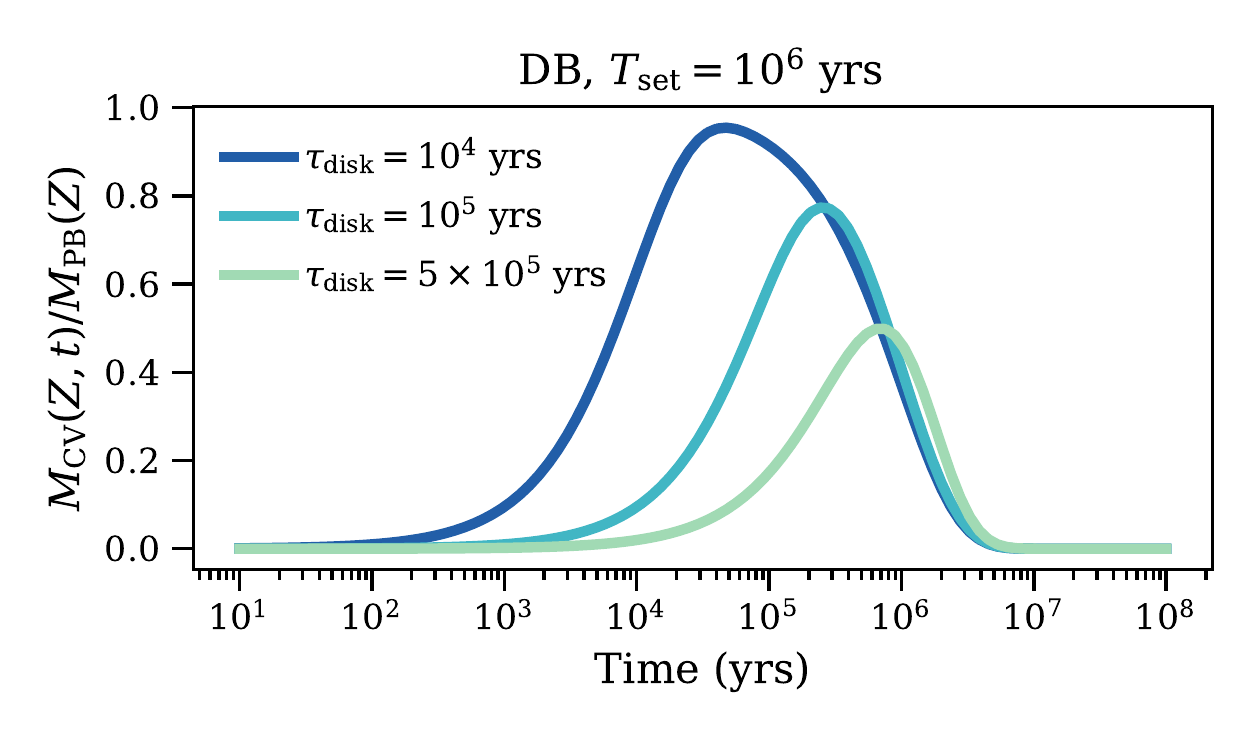}
   \caption{Fraction of parent body mass for a typical rock-forming element in the convection zone of a WD as a function of time, for three different accretion disk lifetimes. The  DA pollution curves decrease by nearly a factor of ten with increasing disk times, while the DB curves decrease less significantly. The peaks of both curves shift to later times as the disk accretion time increases, reflecting  changes in the time where steady state between accretion and settling is achieved.  }\label{MCV_Tdiskvary}
\end{figure}

Estimates for $\tau_{\rm disk}$ generally range from $10^4$ to $10^6$ yr \citep[e.g.,][]{Girven2012}. In Figure \ref{MCV_Tdiskvary} we show how varying the disk timescale changes the pollution curves for the theoretical DA and DB stars shown in Figure \ref{Fig:MCV_example}. In particular, note that the maximum of the DA pollution decreases much more rapidly with increasing disk lifetimes than in the case of DBs. 

We employ J09 in two ways. First, we estimate the parent body masses responsible for observed pollution in a sample of 21 white dwarfs and compare these masses with the distribution of asteroid and moon masses in our own solar system (\textsection \ref{section:MPB}). Secondly, we use the model in conjunction with the average masses of asteroids and moons to determine the timescales of observability for both populations (\textsection \ref{section:observability}). 

\subsection{Steady-State Parent Body Masses}\label{section:MPB}
\begin{figure}
\centering
    \includegraphics[width=0.5\textwidth]{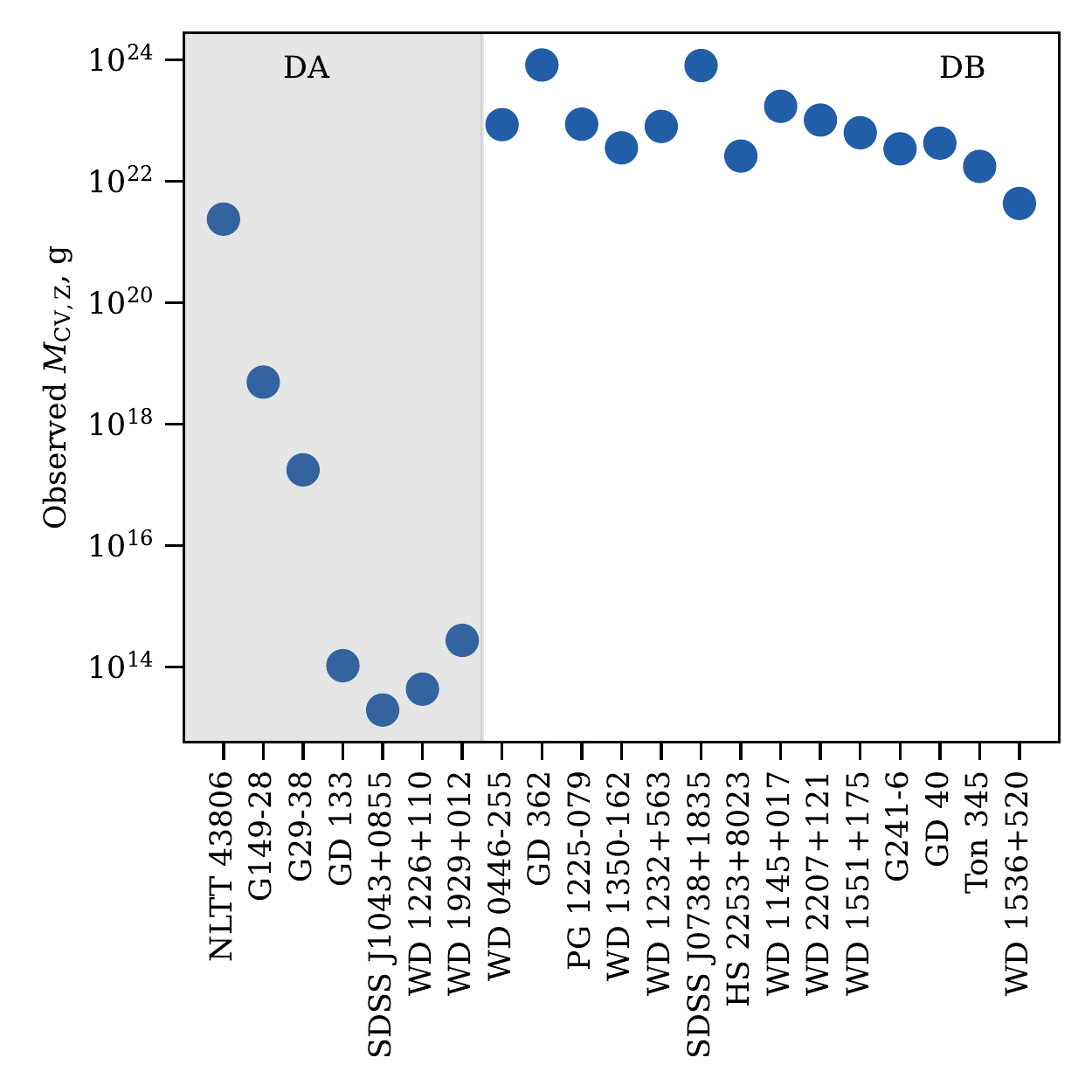} 
   \caption{Total masses of heavy metals observed for the white dwarfs in Table \ref{Table:star_table}, sorted by type. The DAs tend to have lower observed masses of polluting elements compared to the DBs.}\label{Fig:MCVZ_observed}
\end{figure}

In order to calculate parent body masses for polluted white dwarfs, observed pollution masses ($M_{\rm CV}(Z)$) and settling times were collected from the references in Table \ref{Table:star_table}. The total masses of heavy elements in each white dwarfs are shown in Figure \ref{Fig:MCVZ_observed}. For an assumed disk timescale we then solve Equation \ref{Eqn:MCV} for the parent body mass at a range of possible elapsed accretion times for each observed element in the white dwarf. This gives an expression for the mass of element $Z$ in the parent body for an assumed elapsed accretion time $t_{\rm elapse}$ and an observed metal mass of $M_{\rm CV} (Z)$:

\begin{equation}\label{Eqn:MPB_Solution}
    M_{\rm PB} (Z, t_{\rm elapse}) = \frac{M_{\rm CV}(Z) (\tau_{\rm disk} - \tau_{\rm set})}{\tau_{\rm set}\left( e^{-t_{\rm elapse}/\tau_{\rm disk}} - e^{-t_{\rm elapse}/\tau_{\rm set}}\right)}.
\end{equation}

Note that we change the time variable to $t_{\rm elapse}$ to emphasize the difference in the meaning of time between Equations \ref{Eqn:MCV} and \ref{Eqn:MPB_Solution}. Equation \ref{Eqn:MCV} solved for the variation in polluting element mass with time since accretion for a single parent body mass. Equation \ref{Eqn:MPB_Solution} instead takes an observed heavy metal mass and provides a range of parent body solutions that depend on the time at which one assumes the observation was taken.  To obtain the total parent body mass solution, we sum over all observed elements at a given elapsed accretion time, such that $M_{\rm PB} (t_{\rm elapse})= \sum M_{\rm PB} (Z, t_{\rm elapse})$. 

\begin{table*}
\caption{All white dwarf parameters are collected from the references listed in the table. Any values not reported by the paper have been supplemented using the Montreal White Dwarf Database \citep{Dufour2017}. Throughout this work we group white dwarfs by their primary classification type (DA or DB only). }
\begin{center}
\begin{tabular}{c|c|c|c|c|c|c|c|c}
\hline
White Dwarf & Type & $T_{\rm eff}$ (K)  & log(g) & $T_{\rm cool}$ (Gyr) & M$_*$ ($M_\odot$) & log($q$) & Disk$^\dagger$ & Reference \\ \hline 
NLTT 43806 & DAZ & 5830 & 8.00 & 2.4237 & 0.587 & -6.661 & N & \cite{Zuckerman2011}\\
G149-28 & DAZ & 8600 & 8.10 & 1.0683 & 0.657 & -9.224 & N & \cite{Zuckerman2011}\\
G29-38 & DAZ & 11820 & 8.40 & 0.7452 & 0.858 & -12.61 & Y & \cite{Xu2014}\\
GD 133 & DAZ & 12600 & 8.10 & 0.3960 & 0.667 & -15.434 & Y & \cite{Xu2014}\\
SDSS J1043+0855 & DA & 18330 & 8.05 & 0.1095 & 0.649 & -16.645 & Y & \cite{Melis2017}\\
WD 1226+110 & DAZ & 20900 & 8.15 & 0.0812 & 0.714 & -16.663 & Y & \cite{Gansicke2012}\\
WD 1929+012 & DAZ & 21200 & 7.91 & 0.0388 & 0.578 & -16.283 & Y & \cite{Gansicke2012}\\
WD 0446-255 & DBAZ & 10120 & 8.00 & 0.6355 & 0.581 & -5.242 & N & \cite{Swan2019}\\
GD 362 & DB & 10540 & 8.24 & 0.8098 & 0.732 & -5.789 & Y & \cite{Xu2013}\\
PG 1225-079 & DBAZ & 10800 & 8.00 & 0.5343 & 0.582 & -5.235 & Y & \cite{Xu2013}\\
WD 1350-162 & DBAZ & 11640 & 8.02 & 0.4484 & 0.596 & -5.273 & N & \cite{Swan2019}\\
WD 1232+563 & DBZA & 11787 & 8.30 & 0.6623 & 0.773 & -5.924 & N & \cite{Xu2019}\\
SDSS J0738+1835 & DBZA & 13950 & 8.40 & 0.495 & 0.842 & -6.324 & Y & \cite{Dufour2012}\\
HS 2253+8023 & DBAZ & 14400 & 8.40 & 0.4541 & 0.842 & -6.408 & N & \cite{Klein2011}\\
WD 2207+121 & DBZ & 14752 & 7.97 & 0.2046 & 0.572 & -5.591 & Y & \cite{Xu2019} \\
WD 1551+175 & DBAZ & 14756 & 8.02 & 0.2230 & 0.601 & -5.691 & Y & \cite{Xu2019}\\
WD 1145+017 & DBZA & 14500 & 8.11 & 0.2746 & 0.655 & -5.819 & Y & \cite{FortinArchambault2020}\\
GD 40 & DBZA & 15300 & 8.00 & 0.1912 & 0.591 & -5.805 & Y & \cite{Jura2012}\\
G241-6 & DB & 15300 & 8.00 & 0.1912 & 0.591 & -5.805 & N & \cite{Jura2012}\\
Ton 345 & DB & 19780 & 8.18 & 0.1097 & 0.706 & -7.883 & Y & \cite{Wilson2015}\\
WD 1536+520 & DBA & 20800 & 7.96 & 0.0491 & 0.578 & -8.938 & N & \cite{Farihi2016}
\end{tabular} 
\end{center}
{* $q$ is the fraction of stellar mass in the stellar envelope. Note that the DA WDs have much smaller stellar envelopes.\\ $\dagger$ Debris disks indicated for WDs with detected infrared excesses.} \label{Table:star_table} 
\end{table*}

\begin{figure}
\centering
    \includegraphics[width=0.5\textwidth]{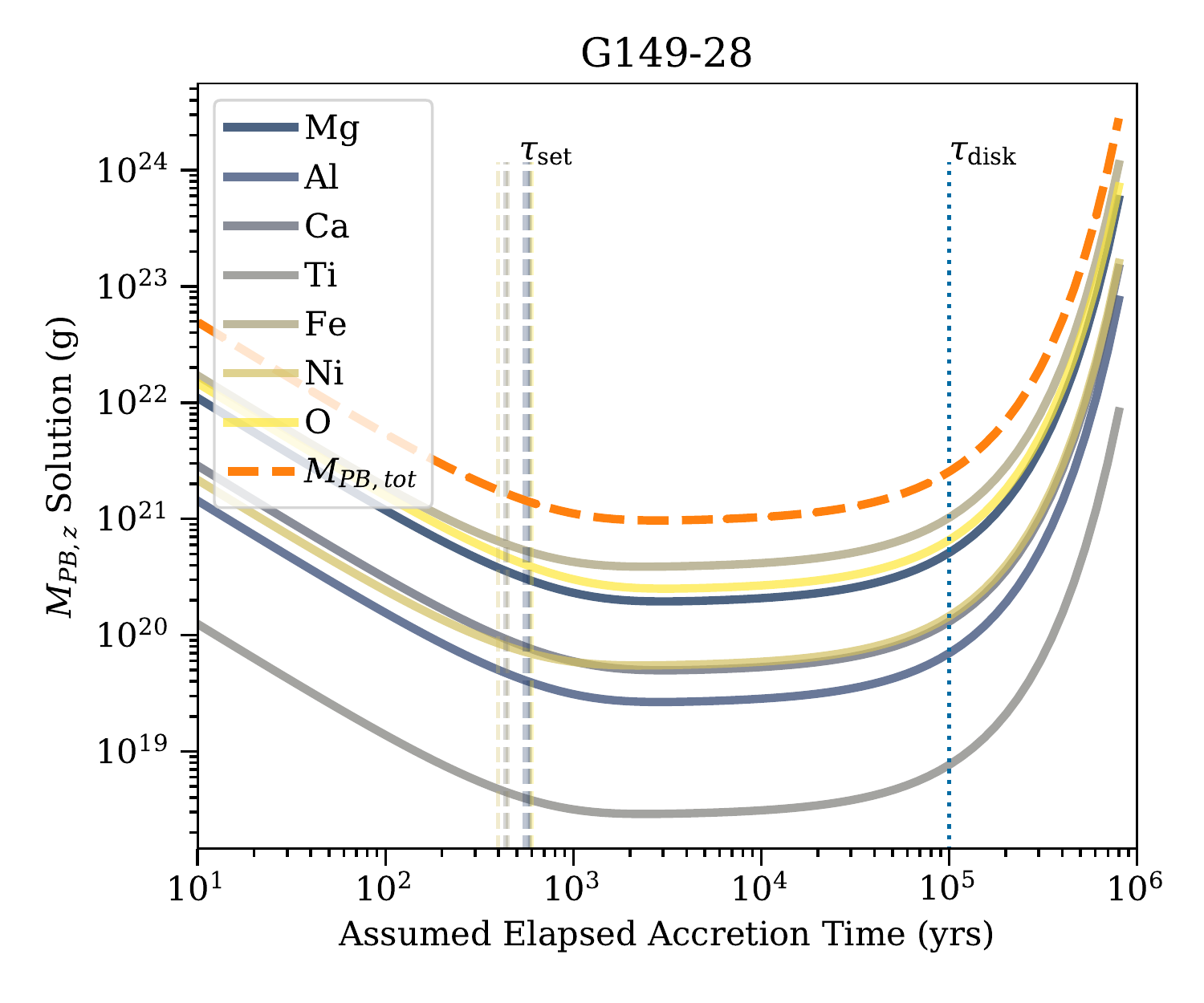} 
    \includegraphics[width=0.5\textwidth]{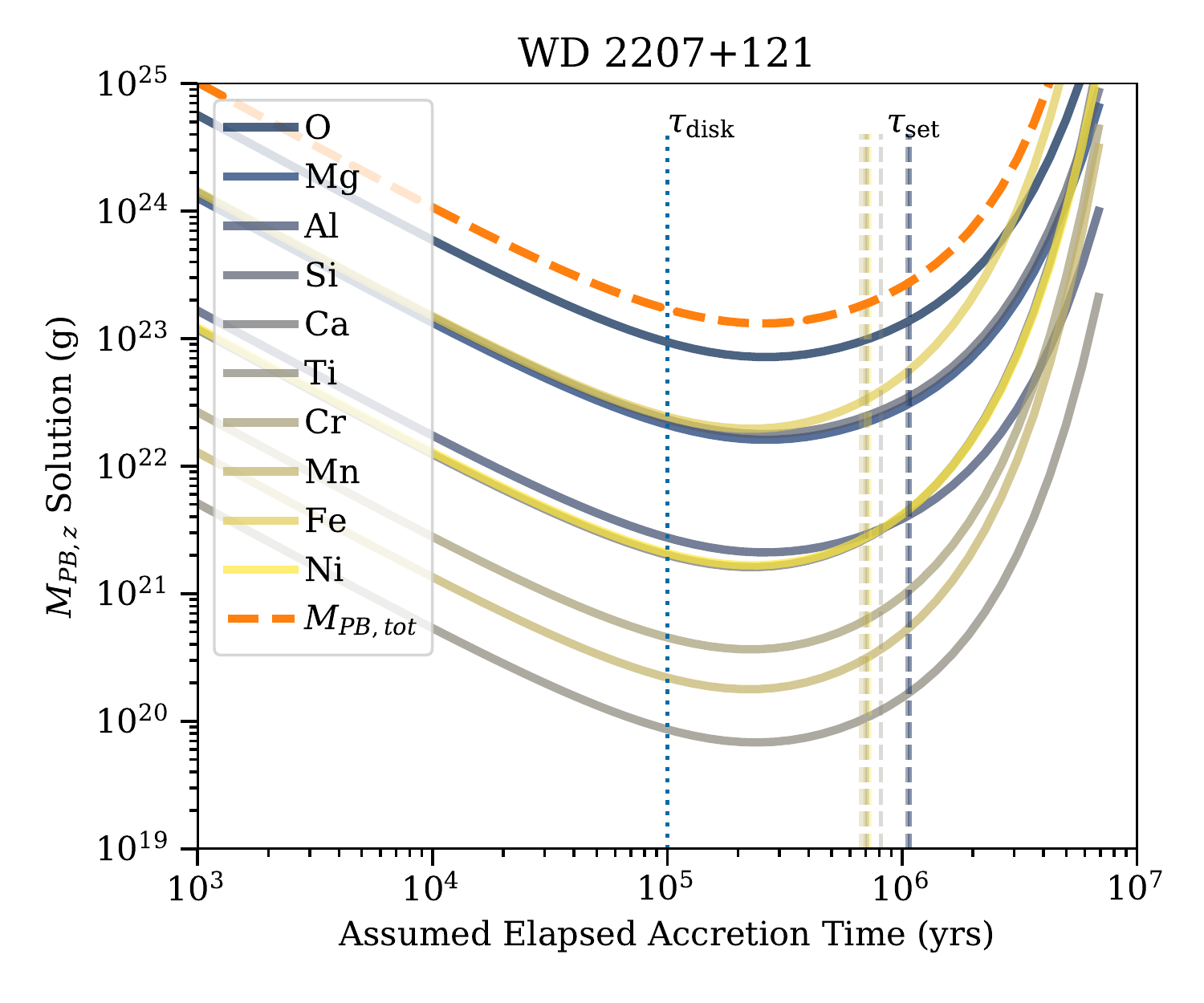} 
   \caption{An example plot of the parent body mass calculated from Equation \ref{Eqn:MPB_Solution} as a function of assumed elapsed accretion time for the DA G149-28 (top) and DB WD 2207+121 (bottom). We assume an accretion disk lifetime of $10^5$ years (dotted line). Settling timescales for the DB are approximately $10^6$ years, those for the DA are $\sim 400$ yr, and both are marked in dashed lines for each element. The parent body solution derived from Equation \ref{Eqn:MCV} is plotted for each element, and the dashed orange line shows the sum of all parent body elements at each assumed elapsed accretion time. }\label{MPB_example}
\end{figure}

As an example, Figure \ref{MPB_example} shows the application of Equation \ref{Eqn:MPB_Solution} to the DA G149-28, and DB WD 2207+121. The vertical lines show the settling times for each element associated with each WD and the assumed disk lifetimes of $10^5 \rm yr$. Note that the shape of the parent body solution is roughly the inverse of the pollution mass curve, reaching a minimum during the steady state phase of accretion, when the pollution mass is at a maximum. Equation \ref{Eqn:MPB_Solution} shows that the steady state point, $\rm dM_{\rm PB} (Z, t_{\rm elapse})/dt_{\rm elapse}=0$, coinciding with the minimum estimate for the parent body mass, will occur at  time:
\begin{equation}\label{Eqn:tmin}
    t_{\rm min} = \frac{\tau_{\rm disk} \tau_{\rm set}}{\tau_{\rm  disk}-\tau_{\rm set}} \ln{\left(\frac{\tau_{\rm disk}}{\tau_{\rm set}}\right)}.
\end{equation}

In practice, when summing over multiple elements as in Figure \ref{MPB_example}, each element would reach the steady state point at slightly different times, due to the variations in settling times. Nonetheless, as long as the range of settling times are well above or below the disk lifetime, solving for elemental abundances at the time corresponding to steady state will give a minimum estimate for the total parent body mass. Thus, for the remainder of this work one can think of the `minimum parent body mass' to be analogous to the `parent body solution assuming steady state.' Note that this does not necessarily mean we are assuming all white dwarfs are in steady state, but rather that any other phase of accretion would require a more massive parent body than the steady state solution to explain the observed metal pollution. 

We calculate parent body masses for the  white dwarfs in Table \ref{Table:star_table} using the effective temperature and log \textit{g} values reported in the references to obtain the WD convection zone masses and elemental settling times from the Montreal White Dwarf Database (MWDD) \citep{Dufour2017}. We then derive $M_{\rm CV}(Z)$ values from the relative abundances reported in the references, using the MWDD settling times and white dwarf envelope mass fractions. For comparison, we also considered the models provided by \cite{Koester2020}, which calculate settling times that include the lack of convection zones in hot, hydrogen dominated white dwarfs. We find that the settling timescales derived in the Koester models are generally comparable to those reported by the MWDD, and therefore do not significantly change the resulting parent body masses. 

Additionally, while we use all stellar parameters from the MWDD instead of those reported in each reference, we find that in most cases the values are in agreement, to within a factor of a few. 

\begin{figure}
\centering
    \includegraphics[width=0.5\textwidth]{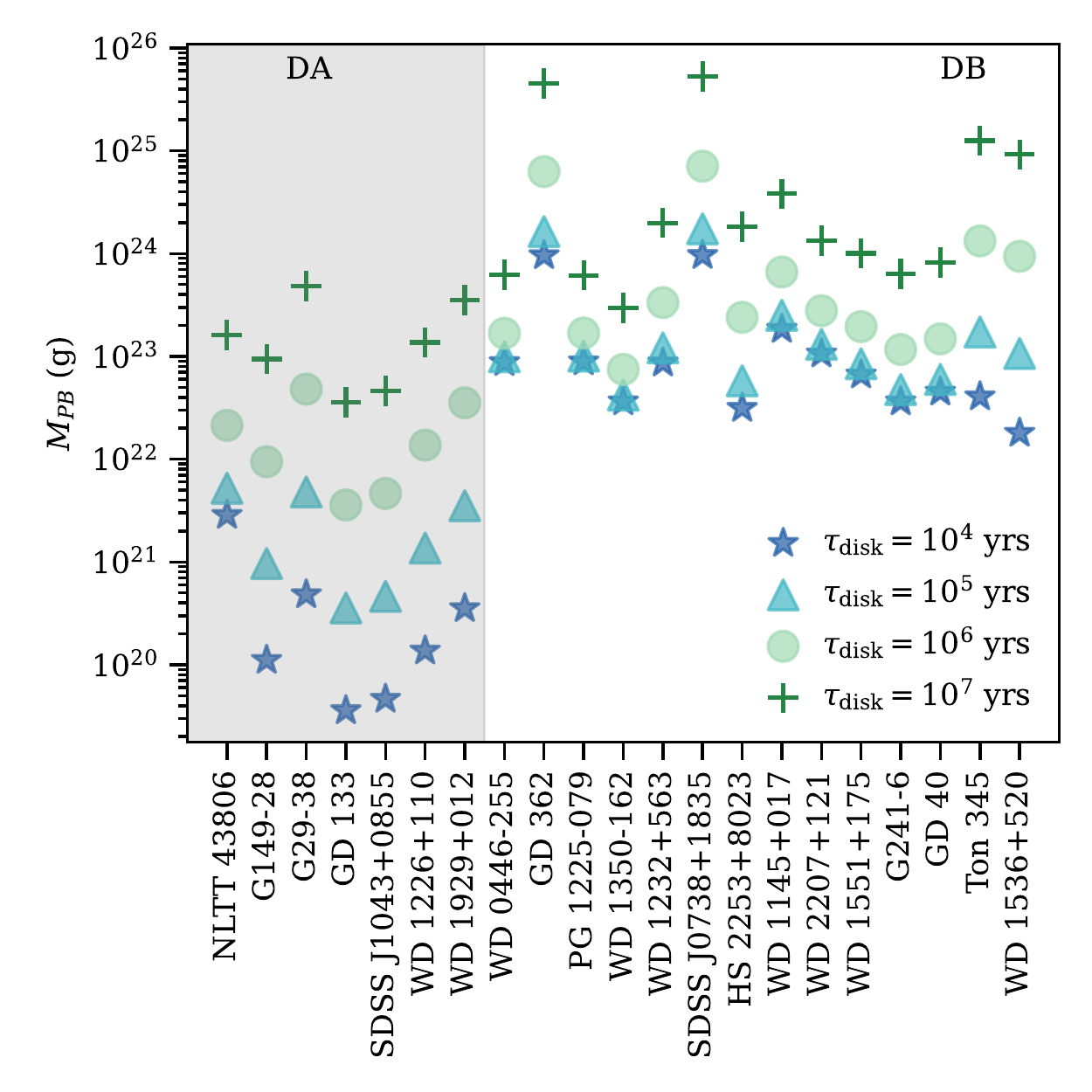}
   \caption{Parent body masses calculated assuming the white dwarf is in steady state, with four different assumed accretion disk timescales. DA parent body solutions tend to vary more dramatically with disk timescale than the DBs. Most DB parent body mass solutions are close to $\sim 10^{23}$,and while the DA solutions are somewhat lower, the two begin to converge as disk timescales increase. Note that the $10^7$ yr timescale is longer than most estimates for disk timescales.}\label{MPB_diskvary}
\end{figure}

Minimum parent body solutions for the observed white dwarfs are shown in Figure \ref{MPB_diskvary}, for a range of disk timescales. Because of the marked dependence on disk lifetime (Figure \ref{MCV_Tdiskvary}), we expect variations in assumed disk lifetime to change the minimum parent body estimates for DAs much more than for DBs. This expectation is realized, as shown in Figure \ref{MPB_diskvary}.

We find that the majority of parent body estimates for DB WDs are between roughly $10^{23}$-$10^{24}$ g regardless of the disk timescale chosen. 
DA estimates are generally lower
unless we increase the disk timescale to $10^7$ yr. This is beyond current estimates for disk lifetimes given in the literature, but does provide
a more satisfactory agreement between parent body masses for DA and
DB white dwarfs. Despite that effect, we will adopt a disk timescale
of $10^5$ yr for the remainder of our analysis as that value is more
generally accepted in the literature and results in lower parent body
masses that we can take as minimum estimates.

\begin{figure}
\centering
    \includegraphics[width=0.5\textwidth]{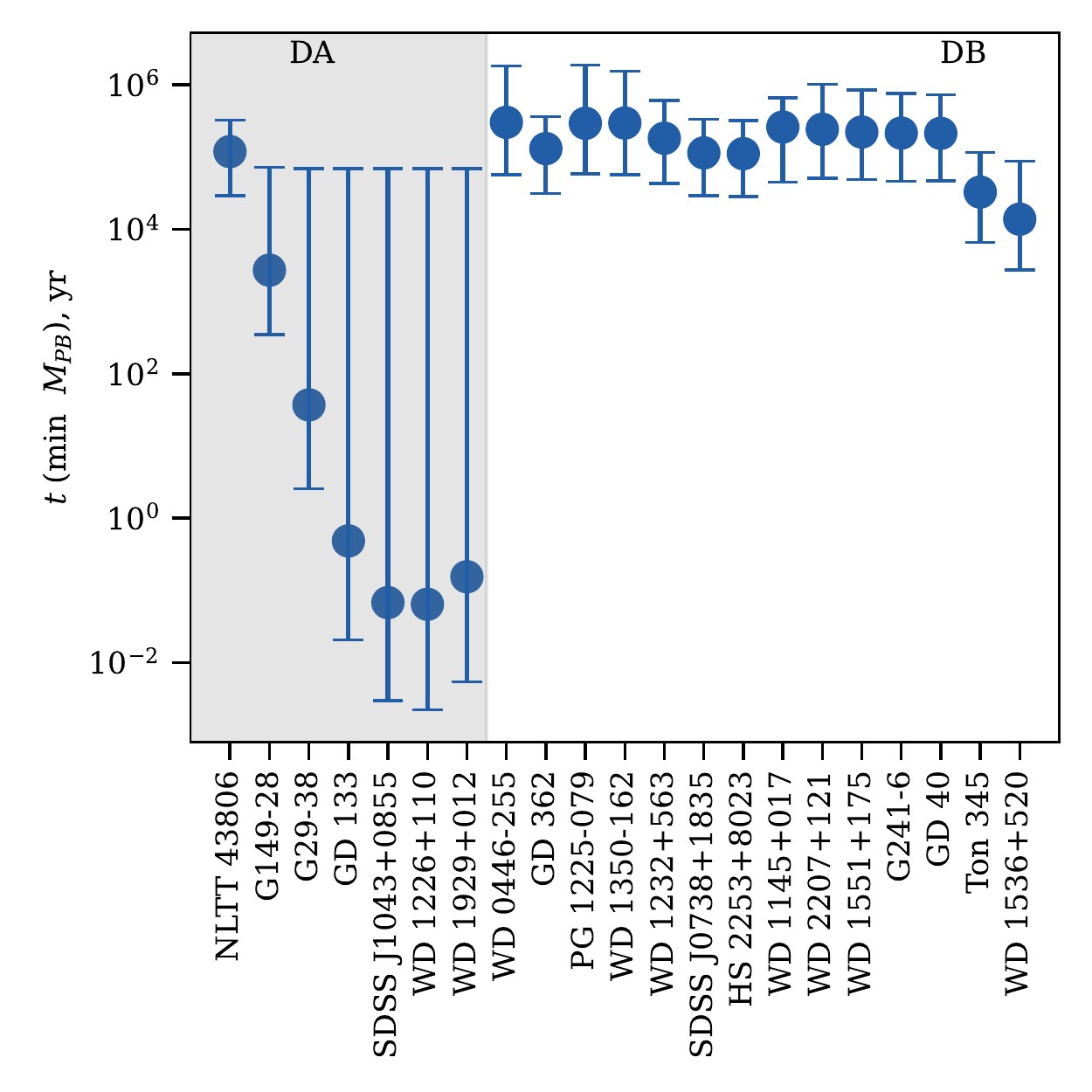} 
   \caption{Elapsed accretion times at which we calculate the minimum parent body mass solution, or equivalently the times where accretion and settling rates are equal. The upper and lower limits for each WD show the range of elapsed accretion times for which the parent body mass is within a factor of two of the minimum. }\label{tmin_sample}
\end{figure}

Figure \ref{tmin_sample} shows the elapsed accretion times corresponding to the minimum parent body solution for each WD in the sample, using the stellar parameters from MWDD, and with a disk lifetime of $10^5$ years. The upper and lower limits show the range of accretion times for which the parent body solution is within a factor of two of the minimum.

\begin{figure}
\centering
    \includegraphics[width=0.5\textwidth]{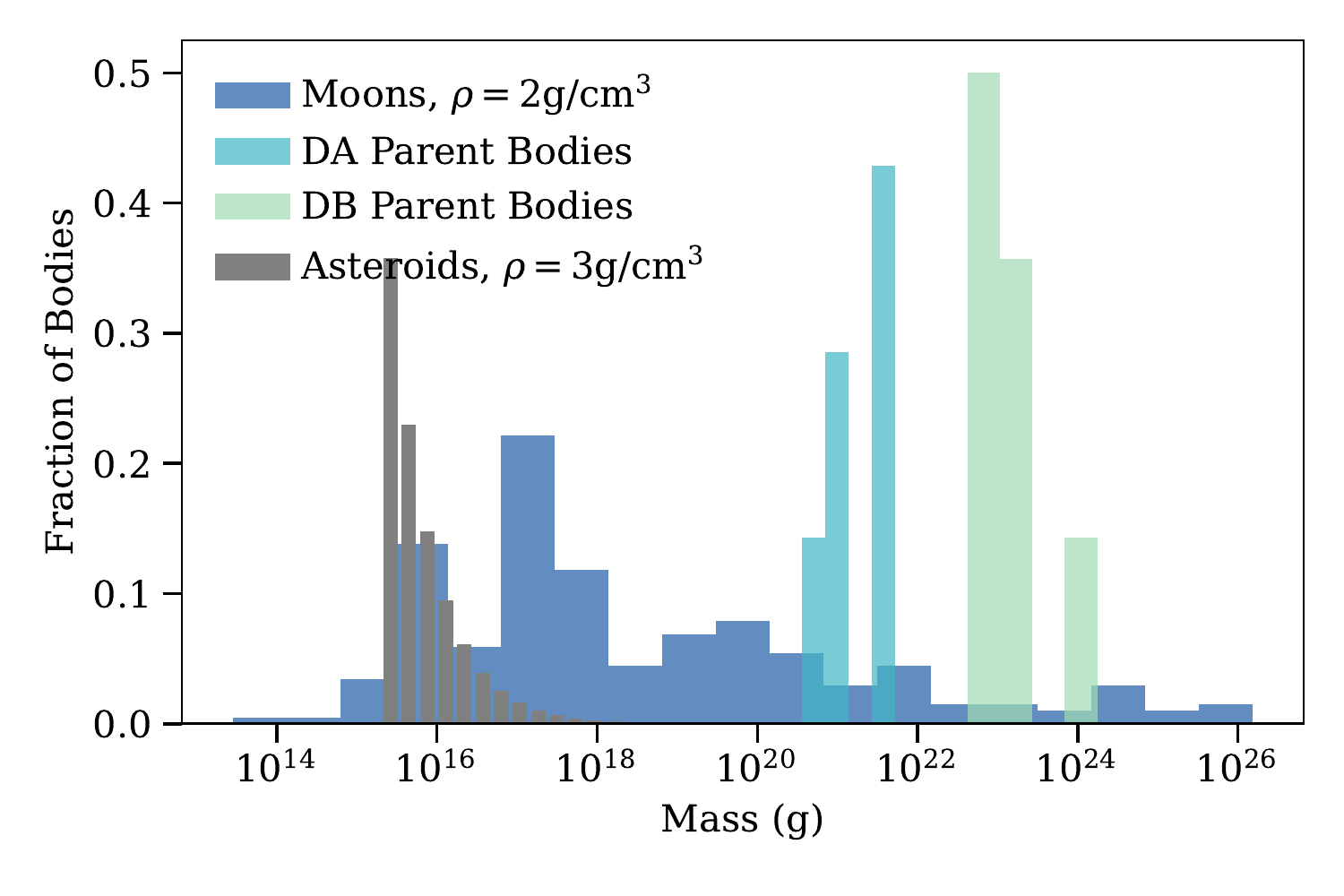}
   \caption{Fraction of bodies in each population with a given mass. Densities of 2.5 and 2 $\rm g/cm^3$ are assumed to calculate masses for asteroids and moons, respectively. The asteroid curve is derived from \cite{Dohnanyi1969} and the moon data are from the JPL Solar System Dynamics group. The parent body masses are split into DA and DB populations and assume a disk lifetime of $10^5$ yr. The DBs tend to have larger parent body mass solutions, though the masses for both the DAs and DBs are situated towards the tail end of asteroid masses and mid-to-high moon masses. Note that the parent body masses shown are lower limit solutions. }\label{MComparison2}
\end{figure}

In Figure \ref{MComparison2} we show how the distribution of calculated parent body masses (assuming a disk lifetime of $10^5$ yr) compares to the distributions of moon masses in the solar system as well as to the approximate distribution of asteroid masses. Moon masses and radii are from the JPL Solar System Dynamics group \footnote{https://ssd.jpl.nasa.gov/ \citep{Giorgini1996} data taken 2021 August}. We calculate the asteroid masses assuming the distribution of asteroid radii is $dN \propto r^{-3.5} dr$ \citep{Dohnanyi1969}, and that all bodies have the same density of 3 g$/$cm$^3$. In reality, the majority of asteroids have lower densities ($\sim 2.5 \rm g/cm^3$), so we can consider the asteroid masses as upper limits. We assume that the range of asteroid diameters is 1-1000 km, corresponding with a range of masses of approximately $10^{15} - 10^{24}$ g. The WD pollution parent body masses are generally far larger than those defined by the asteroid distribution. While the DA parent body masses and the majority of the DB masses reside in the range of the largest bodies in our asteroid belt, such as Ceres ($\sim 10^{24}$ g) or Vesta ($\sim 10^{23}$ g), they are well above the bulk of the asteroid mass distribution. From our sample, for a disk timescale of $10^5$ yr, we consider a mean DA parent body to be  $\sim10^{21}$ g and a mean DB to be $\sim10^{23}$ g. In our solar system, there are about 30 and 15 moons that fall above these DA and DB masses, respectively. 

The large calculated parent body masses compared to minor solar system bodies implicates an observational bias. This may be partially due to the large masses required to  detect pollution, which we outline in the next section. 

\section{Pollution Detection Limits}\label{section:observability}
In \textsection \ref{section:MPB}, we showed that in the context of the J09 model, most observed WD pollution requires parent body masses consistent with the more massive moons of the solar system, and the extremely rare most massive asteroids. 
We now turn to observations to determine a lower limit of observable pollution in a WD atmosphere to place further constraints on the differences between moon and asteroid pollution.

Lower limits for observable masses of pollution in white dwarf atmospheres are obtained from measurements of calcium masses in polluted WDs. We choose calcium because there is a large sample of observations
for Ca available in the literature with which we can assess minimum masses. We collect Ca masses for DA WDs from the SPY Survey \citep{Koester2005} and masses for the DBs from \cite{Zuckerman2010}. We first approximate a line to the lowest calcium masses in the SPY survey to derive an expression for minimum mass as a function of effective temperature (equation \ref{Eq:min_obs_Ca}). We then use the DB data to renormalize the line for the DB WDs. Because the data are expressed by number relative to hydrogen/helium, the relation between the minimum masses and temperature is dependent on the mass of the convection zone (mass of hydrogen for DAs or helium for DBs). Our expressions for the minimum masses of Ca that are observable in the polluted WDs are

\begin{equation}\label{Eq:min_obs_Ca}
\begin{split}
    M_{\mbox{min Ca, DA}} &= 10^{\frac{4 \rm {\it T}_{eff}(K)}{15000} - 12.6} \times \frac{m_{\rm Ca}}{m_{\rm H}}\times M_{\rm CV}\\ 
    M_{\mbox{min Ca, DB}} &= 10^{\frac{4 \rm {\it T}_{eff}(K)}{15000} - 14.2} \times \frac{m_{\rm Ca}}{m_{\rm He}} \times M_{\rm CV},
\end{split}
\end{equation}

\noindent where $m_{\rm Ca}$ is the mass of calcium, $m_{\rm H}$ is the mass of hydrogen, $m_{\rm He}$ is the mass of helium, and $M_{\rm CV}$ is the mass of the WD convection zone or WD atmosphere. As the atmospheres of DAs tend to be relatively small (see Table \ref{Table:star_table}),  their limits of detectable calcium mass are much lower than that of DBs. 

Applying Equation \ref{Eq:min_obs_Ca} to the white dwarfs in Table \ref{Table:star_table}, we obtain the minimum detectable calcium mass in the observable layers for each WD. The DBs have detection limits of $\sim 10^{18}$ g Ca, while the DA limits extend to orders of magnitude lower. The lower limit for calcium detection in the DAs is consistent with their lower observed $M_{\rm CV, Z}$ values, as described above.

\begin{figure}
\centering
    \includegraphics[width=0.5\textwidth]{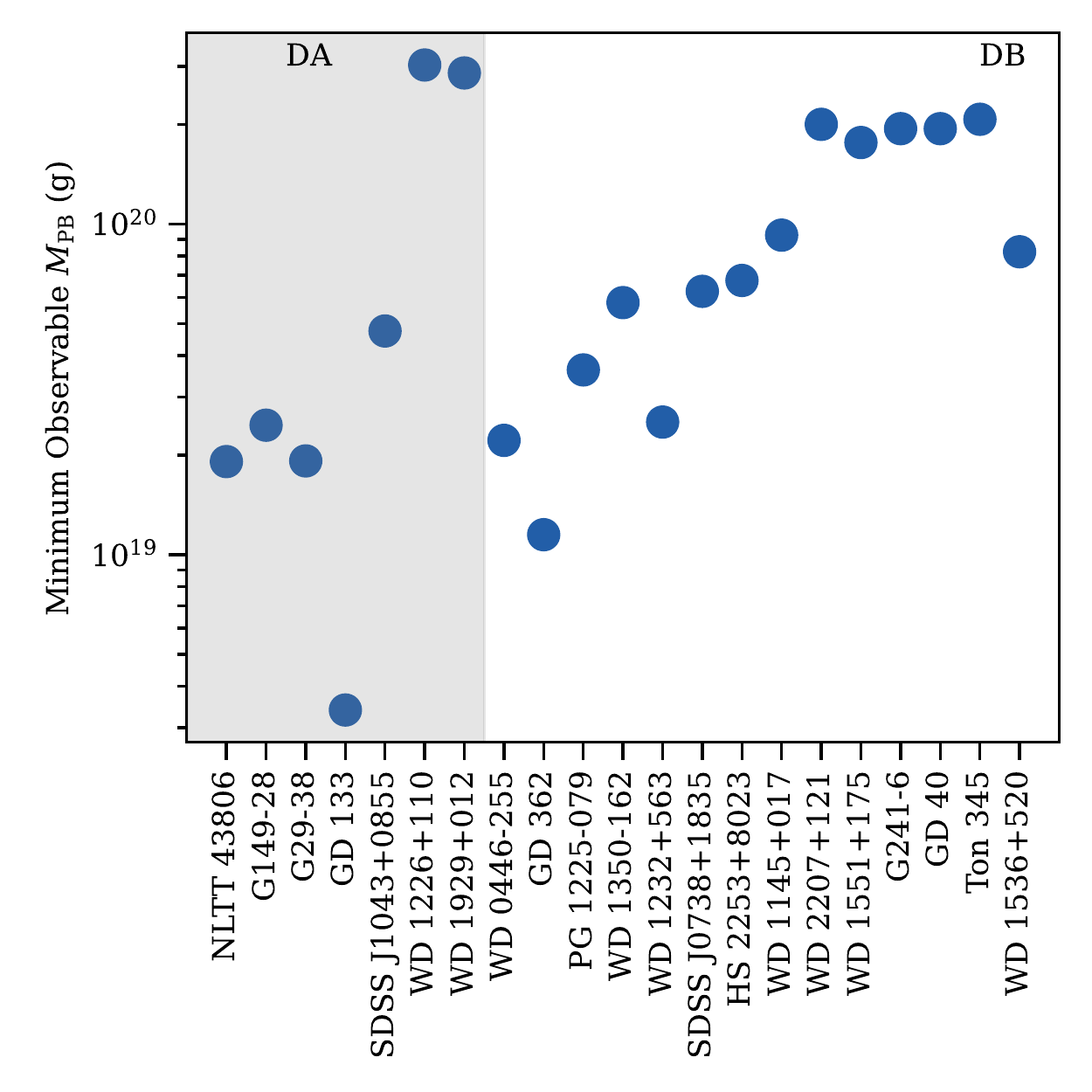}
    \caption{Calculated minimum detectable parent body masses associated with the detection limit for Ca, assuming chondritic  parent body compositions. To obtain these masses, we solve the J09 model for the minimum mass of calcium in the parent body, assuming an $M_{\rm CV, Ca}$ according to Equation \ref{Eq:min_obs_Ca}. We then calculate the minimum parent body mass solution by assuming chondritic composition, settling timescales from MWDD corresponding to the effective temperature of each WD, and a disk timescale of $10^5 \rm yr$. }
    \label{min_obs_PB}
\end{figure}

Taking the minimum observable calcium masses for each of the observed WDs, we now use the J09 model to calculate the parent body mass associated with each calcium mass limit. For each WD, we first apply the J09 model to the minimum detectable calcium mass to find the minimum mass of calcium in the parent body. We then calculate the total parent body mass by assuming chondritic composition ($\sim 1 \%$ calcium by mass), an accretion disk e-folding time of $10^5$ years, and settling times provided by the MWDD based on the effective temperature of each WD. Figure \ref{min_obs_PB} shows the resulting parent body masses associated with the extrapolated minimum observed calcium mass for each WD in our sample. These bodies would provide just enough calcium pollution to be detected with current technology. 

While the instantaneous minimum observable pollution masses derived from equation \ref{Eq:min_obs_Ca} require much higher masses for the DBs, we find that DAs and DBs require similar overall parent body masses to produce observable pollution. This is due to the difference in fractions of parent body that can build up in each type of atmosphere (Figure \ref{Fig:MCV_example}). Assuming each parent body is accreting as a single event, the resulting minimum parent body masses required for observable pollution for both types of WDs are generally larger than the mean for solar system asteroids, and closer to the masses of solar system moons, suggesting an observational bias against ``typical" asteroids. 

In the following sections we will examine how moons and asteroids compare when we allow for continuous accretion of material. This approach is particularly necessary for asteroids, which are thought to reach accretion rates that require material from multiple objects to be present in the WD atmosphere at any one time. Going forward, we will adopt the instantaneous minimum observable mass limits of $5.3 \times 10^{16}$ and $1.3 \times 10^{20}$ g of total heavy elements for a typical DA ($T_{\rm eff} = 10000$ K) and DB ($T_{\rm eff} = 14000$ K),  respectively. For single chondritic accretions, these limits correspond to minimum observable total parent body masses of $1.3 \times 10^{19}$ g for the DA and $1.7 \times 10^{20}$ g for the DB. Note that because DAs accumulate much smaller fractions of parent body in their atmospheres at steady state, the minimum observable parent body masses for DAs and DBs are similar, despite orders of magnitude differences in their instantaneous limits.

\section{Continuous accretion models for asteroids and moons}\label{Section:Asteroids}
While describing the J09 model in \textsection \ref{Jura Model}, we considered the increasing, steady state, and decreasing phases for a single accreting body. However, it is possible that pollution in the WD atmosphere could be from multiple parent bodies. In particular, \cite{Mustill2018} find that asteroids can accrete onto WDs continuously for up to billions of years. To assess how much of the pollution from continuous accretion could fall within observable limits, we consider populations of potential polluters based on mass frequency distributions of asteroids and moons in the solar system as guides. We then use total accretion rates determined previously from N-body simulations to  construct synthetic pollution curves by applying the J09 model (Equation \ref{Eqn:MCV}) to each event.

We currently restrict the comparison of moon and asteroid accretion to a three-planet system of super-Earths and Neptunes, a system that has previously been shown to result in high rates of asteroid accretion by \cite{Mustill2018}. Additionally, we focus on the first 200 Myr past WD formation, the time frame in which asteroid pollution levels are expected to be at their maximum. 

\subsection{Asteroid Accretion}
Mustill's simulations show that in the first few million years after WD formation, a debris belt can reach a peak accretion rate of $\thicksim10^{-4} N_{\rm AB}$/Myr, where $N_{\rm AB}$ is the number of asteroid belt objects. Because the 200 Myr time span is relatively short compared to the full length of time considered in the Mustill simulations, we consider the accretion rate to be approximately constant in our calculations. We therefore use the accretion rate to constrain the total number of accretions that can occur within the 200 Myr period, and then pick the specific accretion event times randomly from a uniform distribution, such that each point in time is equally likely to be the start of an accretion event. 


The masses of accreting bodies for each event are chosen at random, without replacement, from the distribution of masses representing the solar system asteroid belt.  We assume a range of radii of $0.5 - 500$ km and a total mass of the asteroid belt of $3\times10^{24}$ g. \cite{Dohnanyi1969} found the radius distribution for a collisionally-generated debris belt to be $dN \propto r^{-3.5} dr$. Translating this distribution into masses, we have an asteroid mass range of $1.6\times10^{15} - 1.6 \times 10^{24}$ g and $dN \propto \rho^{5/6} m^{-11/6} dm$. The total mass of the belt is then $M_{\rm belt} \propto \int m dN$, which we constrain to be the mass of the asteroid belt. Assuming that all bodies are spherical, we obtain $M_{\rm belt} = A \int \rho \frac{4 \pi}{3}  r^3 r^{-3.5} dr$, where A is a constant, and $\rho$ is the density of the asteroids. Assuming a constant density of 3 $\rm g/cm^3$ and our adopted radius range of $0.5 - 500$ km, we find $dN = 1.7 \times 10^{19} r^{-3.5} dr$. This gives a total of about 12 million objects in the debris belt, for an accretion rate of approximately 1200 events/Myr. We therefore will only accrete a fraction of the total number of bodies in the full 200 Myr time period. 

To sample this distribution, we split the range of asteroid masses into 12 bins, spaced logarithmically in mass, a choice which leaves one object in the Ceres-mass bin. We assign each object in the belt a mass bin according to the distribution described previously, again assuming a constant asteroid density of 3 $\rm g/cm^3$. For each accretion event, we pick an object at random, identify its mass bin, and select a mass at random from the range of masses associated with that bin. We obtain masses of individual elements in the parent body by assuming chondritic composition, and then evolve these masses through the J09 model (Equation \ref{Eqn:MCV}) to track the total masses of polluting elements in the WD atmosphere. The number of bodies in the selected bin is then decreased by one.

\begin{figure*}
\centering
    \includegraphics[width=0.9\textwidth]{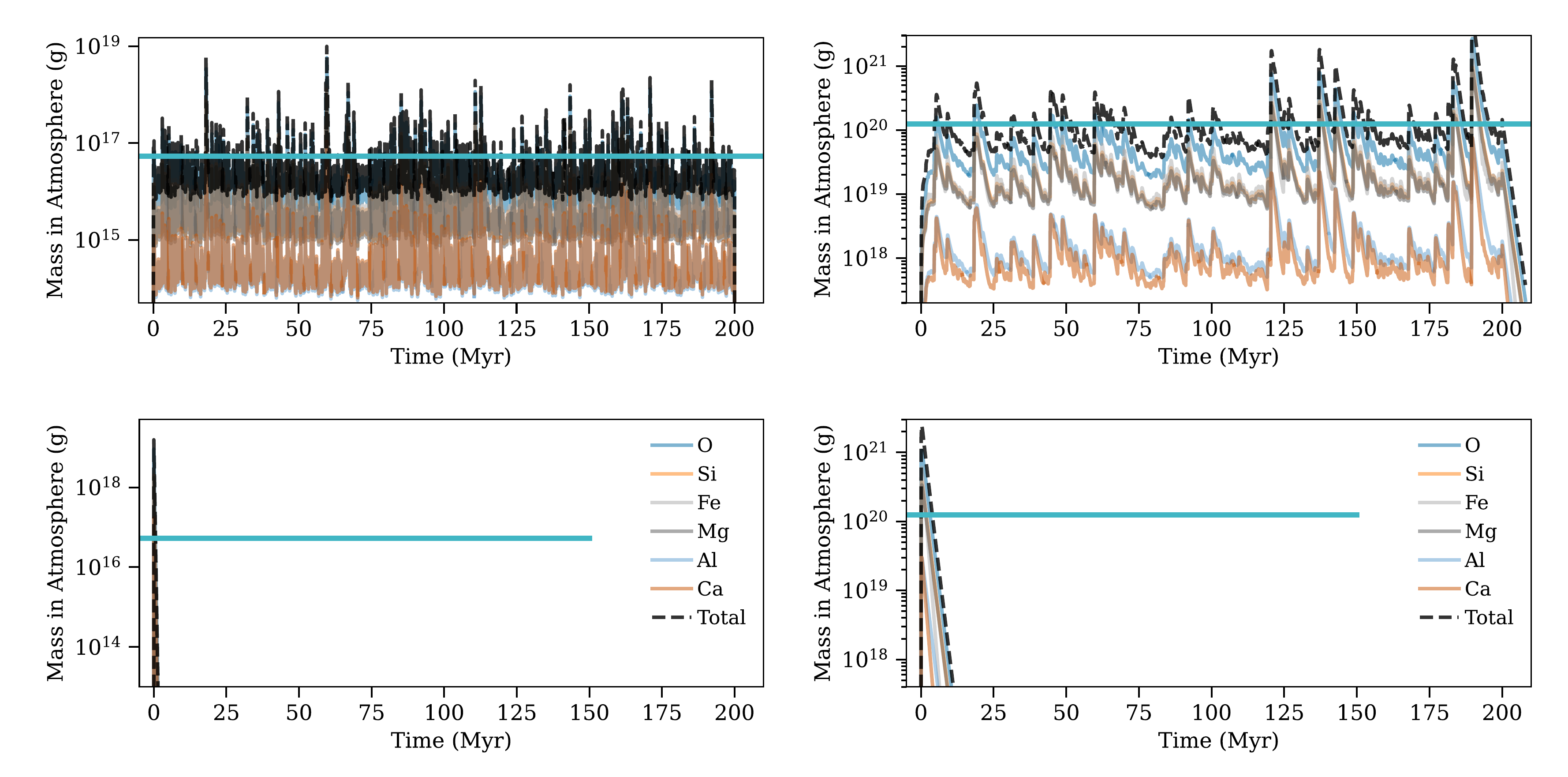}
    \caption{Synthetic pollution mass curve for an asteroid belt (top) and a single moon event (bottom) as a function of time post WD formation. The black dashed line shows the total mass of pollution in the convection zone, the colored lines show the pollution mass per element. The horizontal line shows the minimum detectable mass of heavy elements. The left column shows accretion onto a DA WD while the right shows pollution of a DB.  Top row: Continuous accretion of an asteroid-mass debris belt, based on the mean accretion rate of \cite{Mustill2018} and assuming the distribution of masses follows that of the solar system asteroid belt.
    Bottom row: Accretion of a single moon of mass $\sim10^{20}$ g onto a typical DA (left) and a moon of mass $\sim 10^{21}$ g onto a typical DB (right). These masses represent the median mass of the subset of solar system moons that satisfy the criteria for being liberated from their host planets and providing observable levels of metal pollution.    }\label{fig:asteroid_jura_sim}
\end{figure*}

The upper panels of Figure \ref{fig:asteroid_jura_sim} show the results of this calculation assuming a disk e-folding time of $10^5$ yr and settling times for a 10000 K hydrogen-dominated WD and 14000 K helium-dominated WD.  Each peak in the figure corresponds with an accretion event, and is followed by a tail during which mass sinks out of the atmosphere. The colored curves show the  mass of individual elements comprising the polluting debris in the convection zone as a function of time and the black dashed curve shows the total mass of heavy elements in the convection zone. Note that because asteroid accretions are very frequent compared to the settling timescales, material from at least one body can be found in the mixing layer for the majority of the 200 Myr time interval. 

In \textsection \ref{section:observability} we found that the minimum convection zone pollution mass that is observable is about $5.3\times 10^{16}$ g for a typical DA WD and $1.3 \times 10^{20}$ g for a typical DB, as derived from observed calcium masses. The horizontal lines in Figure \ref{fig:asteroid_jura_sim} show the detectability threshold, and any peaks in  pollution that exceed these limits are considered detectable periods of accretion.

In both the DA and DB cases, a long-term mass of pollution is sustained in the atmosphere, with peaks when more massive asteroids accrete. The sustained background pollution is generally not enough to exceed detection limits, but the more massive asteroid accretions are observable. This suggests that while multiple debris belt objects may be contributing to observed pollution, it is likely that the majority of the observed mass is due to a single, more massive body. We find that asteroid accretions onto the DA can exceed the pollution detection limit for a total of $\sim 29$ Myr, resulting in a cumulative fraction of observable time of $T_{\rm asteroids, DA} = 0.145$

All else equal,  pollution levels are higher for the DB case than the DA case as more mass can build up in the DB WD atmosphere due to slow settling times. Overall, in the DB simulation, the  pollution is at an observable level for a total of $\sim 72$ Myr out of the 200 Myr interval, such that $T_{\rm asteroids, DB} = 0.360$. 

We note that in the asteroid accretion scenarios, the maximum masses of heavy elements that accumulate in the WD atmosphere are $\sim 5\times 10^{16}$ g for the DA and $\sim 10^{20}$ g for the DB. From Figure \ref{Fig:MCVZ_observed}, we see that the WDs in the observed sample have total masses of heavy metals that can exceed the maxima reached by our continuous asteroid accretion simulation by factors of up to $\sim10^3$. In addition to the expected vicissitudes of extrasolar asteroid belt masses,  we consider that the  higher observed masses could be due accretion of moons. 

\subsection{Moon Accretion Simulations}{\label{Section:nbody}}
We carried out the same calculations for moons around  WDs as we did for the asteroids in order to compare the expected detectability of accretion events for the two sources. For this purpose we require an accretion rate for moons.  This rate comes from the efficacy of liberating moons from host planets, and the accretion rate of the liberated moons onto the WD.

We simulate the separation of moons from their host planets during the stellar mass loss event that produces the WDs in order to estimate $f_{\rm moons}$, the frequency of moon accretions by WDs. Because moon orbital periods require very short time steps for integration, we break down $f_{\rm moons}$ into two parts to accommodate computational limits. We define $f_{\rm moons}$ as $N_{\rm moons} \times f_{\rm accrete}$, where $N_{\rm moons}$ is the number of moons in the system that can be liberated from their host planets and are able to provide detectable levels of pollution on a WD and $f_{\rm accrete}$ is how frequently a moon from this population reaches the white dwarf. 

We use the N-body code REBOUND \citep{Rein2012} to model three planets each with two moons.  The IAS15 adaptive time-step integrator \citep{Rein2015} allows time-steps to be shortened or lengthened according to the occurrence of close encounters between particles. Due to computational limits, we set the smallest allowable timestep to be $0.1$ days. Following the approach described in \cite{Payne2017} for moon liberations, we start each simulation with a three-planet system (no moons) and integrate the planet orbits during stellar mass loss. The mass loss excites the planetary orbits, eventually leading to orbit crossings. We halt this portion of the simulation at the first orbit crossing, when the periapsis of any planet falls below the apoapsis of the adjacent interior planet, or, alternatively, when the apoapsis exceeds the periapsis of the adjacent exterior planet. At this point we insert two test particle moons around each planet (6 moons in total). The simulation is then resumed, including stellar mass loss if it is still ongoing. 

During the moon and planet portion of the simulation, we consider a moon to be accreted by the white dwarf if it passes within the Roche limit of the white dwarf ($\sim0.005$ AU). However, we note that it is possible that bodies  farther away than 0.005 AU could still be accreted by the white dwarf, for example through Alfven wave drag \citep{Zhang2021}. Averaging the frequency of moon accretions across all simulation trials gives $f_{\rm accrete}$.

\subsection{Initial Conditions for Moon Simulations}
While Payne et al.\ use planetary architectures as simulated by \cite{Veras2015} and \cite{Veras2016} to assess moon accretion, we carry out our tests using the same three-planet system simulated by \cite{Mustill2018} (\textsection \ref{Section:Asteroids}) in order to compare our results for moon accretions to those of debris belt accretions. 

Each of our simulations begins with three planets around a 3$M_\odot$ main sequence host star that evolves into a $0.75 M_\odot$ white dwarf. The planets have masses of 1.3, 30.6, and 7.8 $M_\oplus$ and initial semi-major axes of 10, 11.6, and 13.07 AU, respectively, as prescribed by the Mustill simulations. We focus on this particular set of planets as they resulted in the largest fractions of asteroids engulfed by the white dwarf in the Mustill study. We choose random initial inclinations in the range $[0^{\circ},1^{\circ}]$,  initial eccentricities of zero, and random values between $0^{\circ}$ and $360^{\circ}$ for all other orbital angles. 

Stellar mass loss is incorporated by updating the stellar mass according to the analytical formulae for single-star evolution by \cite{Hurley2000}. This code calculates stellar properties over 1 Gyr. For simplicity each of our simulations begins at the start of stellar mass loss such that white dwarf formation occurs $\sim100$ Myr after the start of the simulation. 

For the three-planet systems, the first orbit crossing usually occurs before the end of the stellar mass loss event. Because mass loss speeds up significantly towards the end of the 100 Myr interval, the stellar mass is usually still close to $\sim3 M_\odot$, and the semi-major axes of the planets have generally not increased dramatically, when the simulation is paused for moon insertion.

We follow the prescription for moon insertion described by \cite{Payne2017}. The semi-major axis of each moon relative to the planet is chosen randomly in the range $r_{\rm min} < a_{\rm moon}/R_H <0.4$, where $R_H$ is the instantaneous Hill radius of the planet at the moment of the first orbit crossing. We considered two $r_{\rm min}$ values of $0.004$ and $0.04$. This range results in numerically manageable, stable orbits. Payne et al. found that once moons are liberated from their host planet, the initial conditions of the moons cease to matter due to the intense scattering each moon experiences. The initial inclination of each moon is randomly chosen to be within $1^{\circ}$ of the plane of planetary orbits, the eccentricity is set to zero, and all other angles are randomly chosen. 

\begin{figure}
    \centering
    \includegraphics[width=0.5\textwidth]{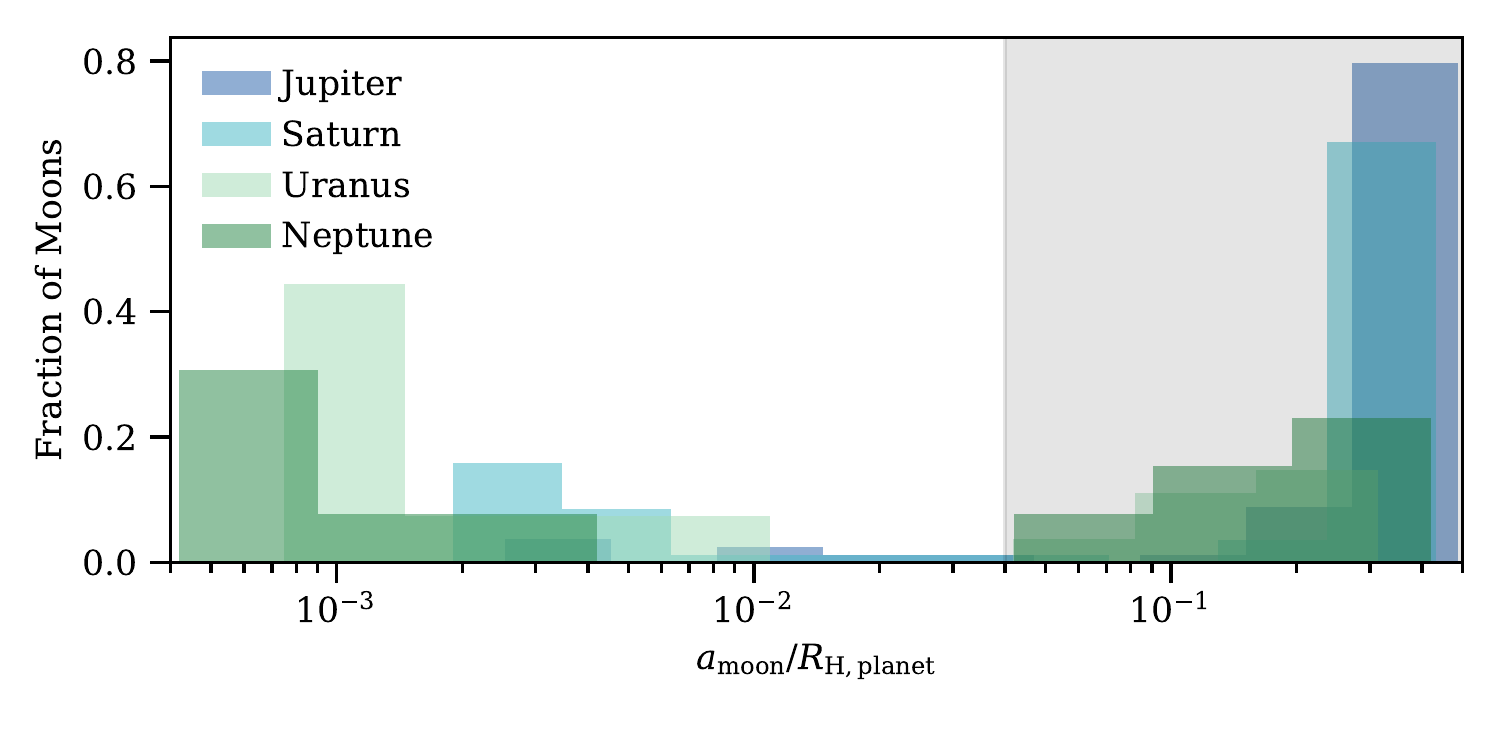}
    \includegraphics[width=0.5\textwidth]{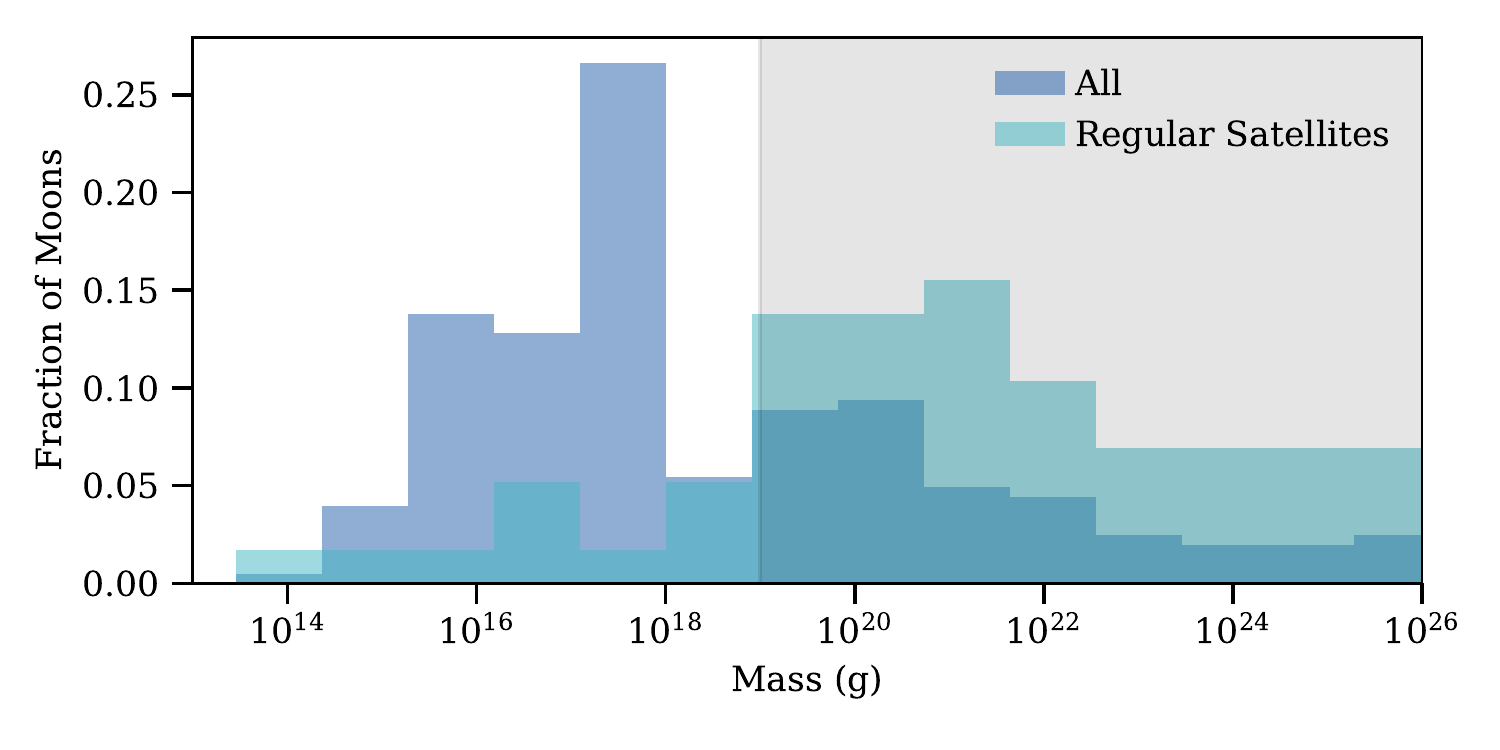}
  \caption{
   Top: Distribution of moon semi-major axes relative to the planet hill radius, for each of the giant planets. 
   Bottom: Distribution of moon masses, including and excluding the irregular satellites. Regular satellites tend to have smaller semi-major axes and larger masses than the irregulars. The shaded portions of each graph show the approximate range of initial semi-major axes that can result in liberated moons (top) and the range of moon masses that can produce observable levels of pollution (bottom), according to our model. }\label{Figure:moon_a_plots}
\end{figure}

For comparison with our initial conditions, Figure \ref{Figure:moon_a_plots} shows the distributions of semi-major axes relative to the host planet Hill radii for solar system moons. Data are gathered from the JPL Solar System Dynamics group, and we assume a uniform density of 2 $\rm g/cm^3$ for all bodies. The top panel shows the distribution of semi-major axes of moons by  planet, and the lower panel shows the masses for the solar system moons. The range in $a_{\rm moon}/R_H$ in our simulations is seen to coincide with the upper third of the values exhibited by solar system moons; semi-major axes of $a/R_H = [0.04, 0.4]$ includes most of Jupiter's and Saturn's moons, but not the majority of Uranus' or Neptune's moons (Figure \ref{Figure:moon_a_plots}).

\subsection{Accretion Rates of Liberated Moons}
\begin{figure*}
    \centering
    \includegraphics[width=0.70\textwidth]{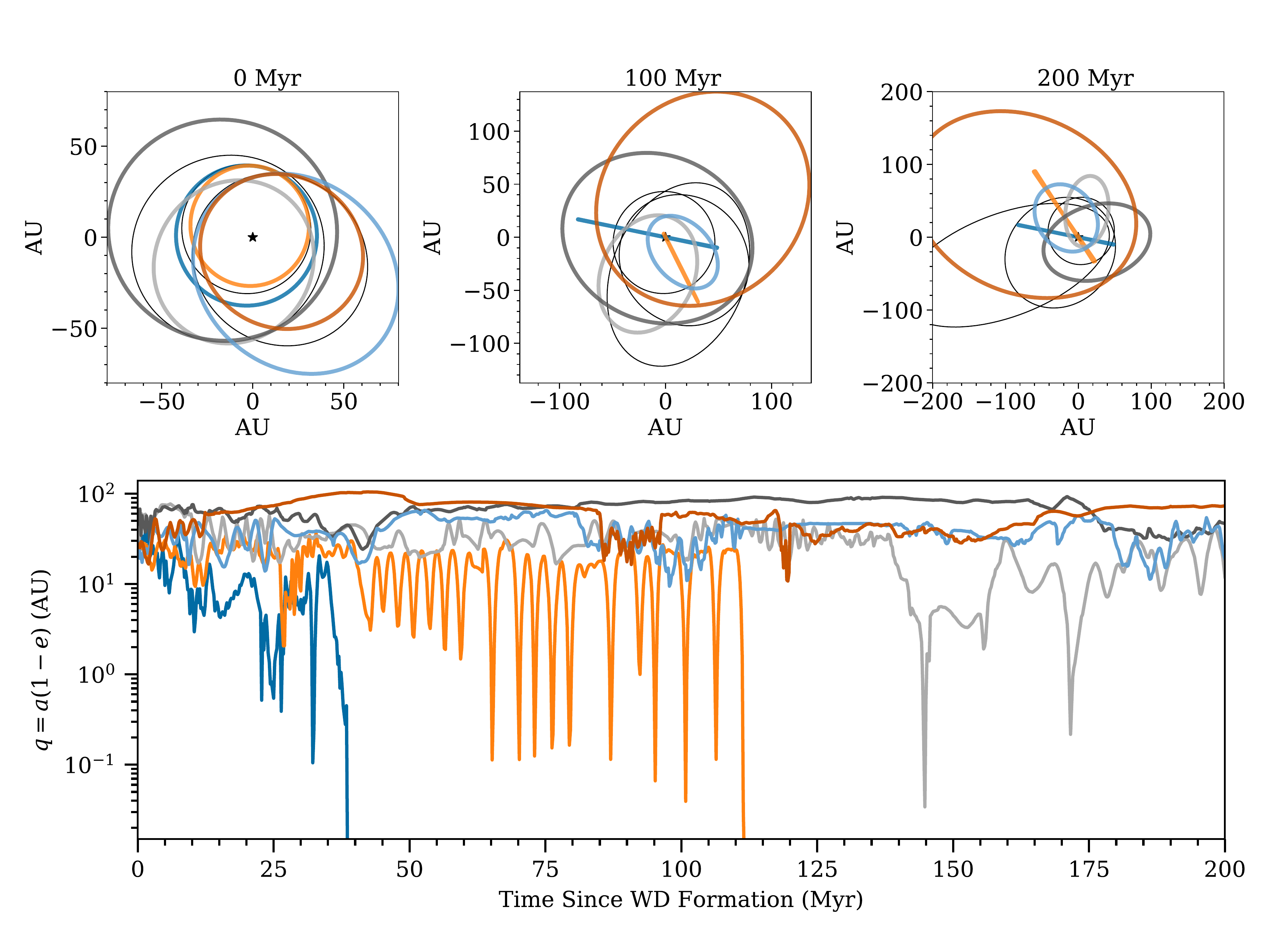}
   \caption{Top: Snapshots of orbits from simulation, for planets (thin black lines) and moons (colors). The times indicate the time since the WD formed, and the WD is shown as a black star at 0 AU. 
   Bottom: Periapse versus time for the moons, in the same simulation as show in the top plot. Each curve corresponds with the orbits of the same color in the top plot. Two moons are scattered onto hyperbolic orbits following a close approach to the WD.  }
   \label{Figure:summary_q}
\end{figure*}

 Figure \ref{Figure:summary_q} shows an example of the results from one simulation. The lower plot shows the instantaneous periapse of each moon, while the upper figures show snapshots of the orbital configurations. In this simulation, the closest approach of a moon to the WD is $\sim 0.007$ AU. Table \ref{Table:nbody} lists the closest approaches of each of the moon test particles across all simulations, as well as the initial conditions for each moon's orbit, relative to its host planet.

Of the 60 moons comprising 10 simulations, one enters the Roche limit 
($\lesssim$0.005 AU) of the white dwarf within the first 200 Myr after white dwarf formation. We therefore set the accretion rate of liberated moons at 1/60 per 200 Myr after the formation of the host WD. Note that like the asteroid N-body results, the moon accretion rates are dependent on the number of bodies available in the system.  Our result should be taken as the fraction of liberated moons that accrete.

In our simulations, all moons are liberated from their host planets. Consistent with \cite{Payne2017}, as a conservative estimate, we assume that any objects outside of $0.04 R_{\rm H}$ of their planet will be liberated. We therefore take the population of solar system moons with semi-major axes in this range as the population of potential moon parent bodies that could pollute a WD. In our solar system, out of a total of about 200 moons, 145 are situated at more than 0.04 $R_{\rm H}$, including the irregular satellites. 

Because moon accretions are single events, we can use the J09 accretion model to set a lower limit on the mass of a moon that can provide the minimum observable mass. For the DA of 10000 K, the least massive observable parent body is $1.3 \times10^{19}$ g and for the 14000 K DB the limit is $1.7\times10^{20}$ g. For the population of solar system moons exterior to our limit for liberation of $0.04 R_{\rm H}$, this gives a total of 22 moons that can provide observable pollution on a DA and 10 that can do so on a DB.

We now return to our expression for the frequency of moon accretions, $f_{\rm moons}$ as $N_{\rm moons} \times f_{\rm accrete}$. We found that 1/60 of moons liberated from their host planets are expected to find their way to the WD. Inserting the accretion rate from the N-body simulations, and considering the number of moons that can both be accreted and observed on the surface of the WD based on the solar system population of moons, gives $f_{\rm moons, DA} = 22 \times 0.017/200 \rm Myr = 0.0019 \rm Myr^{-1}$ and $f_{\rm moons, DB} = 10 \times 0.017/200 \rm Myr = 0.0009 \rm Myr^{-1}$. 

\begin{table*}
\caption{Minimum periapse ($q$) reached by each moon, in AU, and the parameters used to initialize the moon orbits. $a_{\rm moon}$ is given relative to the Hill radius of the planet. Two moons were inserted around each planet. The WD Roche limit is at 0.005 AU.}\label{Table:nbody}
\begin{tabular}{c|c|c|c|c|c|c|c}
\hline
Run & $a_{\rm moon}/R_{\rm H}$ range  & $q1_{\rm min}$ (AU) & $q2_{\rm min}$ & $q3_{\rm min}$ & $q4_{\rm min}$ & $q5_{\rm min}$ & $q6_{\rm min}$ \\ \hline
1 & 0.004-0.4 &  0.0612 & 0.0148 & 0.0060 & 20.1982 & 0.0018 & 0.0568   \\
2 & 0.04-0.4 &  0.6077 & 4.7049 & 0.0619 & 6.6298 & 7.5362 & 2.4578   \\
3 & 0.004-0.4 &  0.2180 & 0.1425 & 0.4133 & 1.5700 & 0.0328 & 0.8134   \\
4 & 0.004-0.4 &  0.0078 & 0.0081 & 7.9765 & 0.6881 & 0.0187 & 0.0066   \\
5 & 0.004-0.4 &  0.9560 & 1.8188 & 0.8765 & 0.0097 & 3.7238 & 0.0070   \\
6 & 0.04-0.4 &  23.2735 & 22.1167 & 15.3154 & 4.8946 & 18.0538 & 17.4935   \\
7 & 0.04-0.4 &  19.7743 & 33.2792 & 28.4050 & 19.1263 & 7.7563 & 31.6862   \\
8 & 0.04-0.4 &  15.2427 & 11.6404 & 8.7045 & 20.0703 & 7.2406 & 26.1100  \\
9 & 0.04-0.4 &  0.0069 & 0.0071 & 0.0341 & 24.2026 & 9.4522 & 10.7774 \\
10 & 0.04-0.4 &  8.4775 & 21.0213 & 8.5416 & 0.0065 & 44.6118 & 27.5651 
\end{tabular}
\end{table*}

\subsection{J09 Model for Moons}
Applying the accretion rates of $f_{\rm moons, DA}= 0.0019 \rm Myr^{-1}$ and $f_{\rm moons, DB} = 0.0009 \rm Myr^{-1}$ results in 0.38 and 0.17 accretions in the first 200 Myr past WD formation, for the DA and DB WDs, respectively. One concludes that moons are expected to be visible as single accretion events, with no build up of pollution from multiple objects, unlike the case for asteroids. 

We use solar system moons as a guide for computing the median mass expected for accretion events with moons as parent bodies. In the previous section, we showed that only moons with total masses greater than $1.3\times10^{19}$ g and $1.7 \times 10^{20}$ g can be detected on the surface of a DA and DB, respectively. In order to obtain the population of observable moons, we therefore select the solar system moons that are above these mass limits, and are situated outside of our assumed liberation limit of $0.04 R_{\rm H}$. We assume the resulting moons represent the population of bodies that could both be liberated from their host planets and provide a detectable amount of pollution if they were to be accreted by their host WD. Recall that we found that 22 solar system moons meet these requirements for the DA accretion events and 10 solar system moons for the DB accretion events. Assuming any of these moons would be equally likely to accrete, we used the median mass moon of each population to represent the median  mass moon expected to be observed accreting onto each WD type. This results in a median mass for an observed accreted moon on a DA of $1.2 \times 10^{20}$ g and $3.7 \times 10^{21}$ g for a DB. 

In the lower panels of Figure \ref{fig:asteroid_jura_sim}, we show the pollution curve due to accretions of these median mass objects on the same time axis as the asteroid accretions to illustrate how the single events compare with the continuous  accretions. We assume that each moon has chondritic composition. Because these are single events, they follow the same phases of accretion as shown in Figure \ref{Fig:MCV_example}, however we now also show the variation in the abundances of individual elements. The heavy metal limit outlined in \textsection \ref{section:observability} is shown as a horizontal line in each plot. 

For a single moon accretion, the DA case exceeds the mass limit for $0.57 \rm Myr$ while the DB is observable for $3.89 \rm Myr$. Note that these timescales far exceed the duration of observability for a single asteroid on either WD type. However, because asteroid pollution accumulates from multiple bodies, the continuous asteroid calculations result in greater cumulative timescales of observability.

For a mean fraction of observable time for the moons we multiply these numbers by the expected number of moon accretions in the 200 Myr time period for each case, obtaining $T_{\rm moons, DA} = 0.38 \mbox{ accretions} \times 0.57 \rm Myr / 200 \rm Myr = 0.001$ and $T_{\rm moons, DB} = 0.17 \mbox{ accretions} \times 3.89 \rm Myr / 200 \rm Myr = 0.003$.

We can now use our results to evaluate the relative probabilities of detecting asteroids from a debris belt and moons in polluted WDs.  

Returning to Equation \ref{Eqn:Pmoons}, we now fill in the values derived for the DA and DB cases to find the relative probability of observing accretion of moons and asteroids, yielding
\begin{equation}
        \left( \frac{P_{\mbox{moon accretion}}}{P_{\mbox{asteroid accretion}}}\right)_{\rm DA} = \frac{0.001}{0.145} = 0.007
\end{equation}
and

\begin{equation}
          \left( \frac{P_{\mbox{moon accretion}}}{P_{\mbox{asteroid accretion}}}\right)_{\rm DB} = \frac{0.003}{0.360}=0.008. 
\end{equation}
Therefore, for three-planet Super-Earth/Neptune systems with both moons and asteroids available for accretion, we would only expect up to $1\%$ of polluted DAs and DBs to currently have observable amounts of pollution due to moon accretion. 

\section{Discussion}\label{Section:Summary}
Over 1000 white dwarfs have observations of at least one polluting element in the atmosphere \citep{Coutu2019}. Around 20 are considered strongly polluted, with multiple rock-forming elements detected. Therefore, $\sim2\%$ of all polluted WDs can be considered `highly polluted'. The large, moon-like mass solutions calculated in \textsection\ref{Jura Model} represent the parent bodies associated with this highly polluted sample. Assuming that all of these high-mass parent bodies are indeed moons, this suggests that $\sim2\%$ of polluted white dwarfs are accreting moons. Of course, as shown by the pollution masses in Figure \ref{fig:asteroid_jura_sim}, accretion of the most massive asteroids may also result in high masses of pollution, so the population of the most highly polluted white dwarfs may also include the most massive debris belt members. Nonetheless, this statistic is consistent with our results derived from N-body accretion rates, that $\sim1\%$ of overall pollution is expected to come from moons as opposed to less massive debris belt objects. 

Uncertainties in this study include the timescales assumed in the J09 model, observational constraints beyond what has been considered in our minimum detectable mass method, exoplanet moon and asteroid populations, and the effects of planetary architectures beyond the three-planet system considered as our test case. We now explore these various possibilities and their effects on the parent body solutions or numerical accretion models.

\subsection{Disk e-folding timescale}
The e-folding lifetime of the debris disks around the WDs, $\tau_{\rm disk}$, determines how quickly pollution accretes onto the WD and, in conjunction with the settling times, limits the maximum mass of any given element that can accumulate in the WD atmosphere (Figure \ref{MCV_Tdiskvary}). Throughout the asteroid and moon comparison in this paper we assume $\tau_{\rm disk} = 10^5 \rm yr$, as an accommodation for estimates that span from $10^4$ to $10^6 \rm yr$. 

\begin{figure}
\centering
    \includegraphics[width=0.5\textwidth]{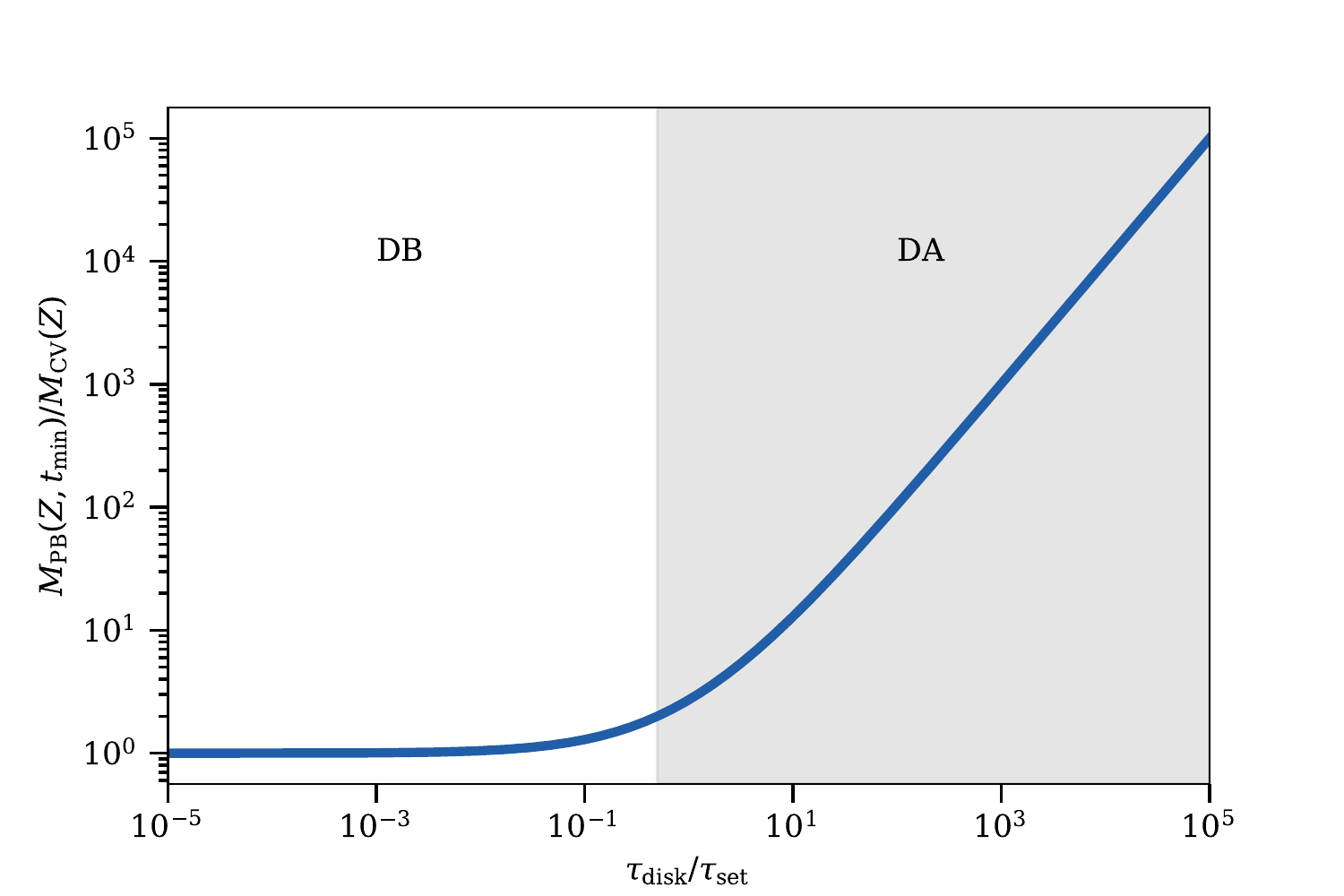}
    \caption{Minimum parent body mass solution relative to the observed mass of pollution in the atmosphere, as a function of the ratio of the disk timescale to the settling time. DAs would be found towards the right, so that the steady-state parent body mass is several times the observed mass in the atmosphere. DBs would be towards the left, where the parent body solution approaches the mass of the polluting metals. }\label{fig:PBvtauratio}
\end{figure}

In \textsection\ref{Jura Model} we derived $t_{\rm min}$, the assumed elapsed accretion time which recovers a minimum solution for the parent body mass, or equivalently the steady state point in the J09 model. Plugging in $t_{\rm min}$ to the J09 model, we can therefore write the general solution for the minimum parent body mass solution relative to the observed mass in the atmosphere as a function of the ratio of the disk and settling timescales, for an arbitrary element:

\begin{equation}\label{Eq:MPBminratio}
    \frac{M_{\rm PB}(Z, t_{\rm min})}{M_{\rm CV}(Z)} = \frac{\frac{\tau_{\rm disk}}{\tau_{\rm set}} -1}{\left(\frac{\tau_{\rm disk}}{\tau_{\rm set}}\right)^{\frac{1}{1 - \tau_{\rm disk}/\tau_{\rm set}}} - \left(\frac{\tau_{\rm disk}}{\tau_{\rm set}}\right)^{\frac{1}{\tau_{\rm set}/\tau_{\rm disk} -1}}},
\end{equation}
where $M_{\rm PB}(Z, t_{\rm min})$ is the parent body solution assuming steady state for the element $Z$ and $M_{\rm CV}(Z)$ is the observed mass of the heavy metal. Figure \ref{fig:PBvtauratio} shows Equation \ref{Eq:MPBminratio} applied to a range of disk-to-settling time ratios. 

Writing the parent body mass expression in terms of the ratio of characteristic timescales shows why DAs and DBs have different sensitivities to changes in $\tau_{\rm disk}$ (outlined in \textsection\ref{Jura Model}) by illustrating the two limits of the parent body mass calculation. If disk timescales far exceed settling times, as they would for DAs, the minimum parent body solution will increase rapidly. For $\tau_{\rm disk} = 10^5$ yr and a DA settling time of days, parent body solutions can reach factors of $10^7$ times the observed metal mass. On the other hand, if settling times exceed disk timescales, as they may for DBs,  $\tau_{\rm disk}/\tau_{\rm set}$ approaches 1, so that the minimum parent body mass is equal to current mass of metal in the WD atmosphere. 

Increasing the assumed disk timescale would shift parent body mass solutions to the right in Figure \ref{fig:PBvtauratio}. While DB WDs with very long settling times would not be strongly affected (parent body solutions would still be roughly equivalent to the mass of metal in the atmosphere), minimum parent body solutions trend approximately linearly with $\tau_{\rm disk}/\tau_{\rm set}$ when the disk timescale exceeds the settling time (DAs). Decreasing the assumed disk timescale would similarly not strongly affect the DBs but would decrease the parent body solutions of the DAs proportionally. Note that WDs with settling times within the range of disk timescale estimates will have non-linear dependencies on disk timescale. These effects can be seen in the parent body solutions calculated for the observed WDs (Figure \ref{MPB_diskvary}).

We now return to our parameters $T_{\rm moons}$ and $T_{\rm asteroids}$, the timescales during which pollution exceeds detectable levels. For simplicity, we start by considering a single accretion with a generic observability timescale $T$. For a single event, $T$ depends on the peak mass of pollution ($M_{\rm CV}$) that can build up in the atmosphere, the duration for which a high mass of pollution can be sustained, and the detection limit associated with the WD. 

The maximum mass of pollution deposited by a given parent body mass is the inverse of Equation \ref{Eq:MPBminratio}. Therefore, pollution accumulation for DAs varies inversely with changing disk timescales while peak masses of pollution in DBs remain roughly constant. The amount of time that relatively large masses of pollution can be sustained in the WD atmosphere can be approximated as the difference between $\tau_{\rm disk}$ and $\tau_{\rm set}$. As seen in Figure \ref{MPB_example}, these timescales are on either side of $t_{\rm min}$, where the maximum $M_{\rm CV}$ is reached. For simplicity, we will consider DA settling times well below, and DB settling times well above, the range of possible disk timescales, so that $|\tau_{\rm disk} - \tau_{\rm set}|$ will be approximately equal to the longest of the two timescales. 

Based on this approximation, for DBs $|\tau_{\rm disk} - \tau_{\rm set}|$  $\sim \tau_{\rm set}$, and to a reasonable approximation, changing $\tau_{\rm disk}$ should not affect $T$ for the DBs. Therefore, and we anticipate $T_{\rm moons}$ and $T_{\rm asteroids}$ to be robust against  disk timescales for DBs of sufficiently long settling times. 

DA settling times are short, so $|\tau_{\rm disk} - \tau_{\rm set}| \sim \tau_{\rm disk}$, and increasing the disk timescale will lengthen the amount of time that peak pollution levels can be sustained. However, increasing $\tau_{\rm disk}$ decreases the maximum mass of pollution that can be accumulated. If the decreased pollution masses still exceed the detection limit, then $T$ would increase, but if the new pollution masses do not exceed detection limits, $T$ would fall to zero. The overall effect on $T_{\rm moons}$ and $T_{\rm asteroids}$ will therefore depend on the detection limits and distribution of debris belt masses considered. For our distributions of moons and asteroids, generally only the most massive bodies are contributing towards observable pollution, so assuming that most pollution would remain above the detection threshold, we would expect $T_{\rm moons}$ and $T_{\rm asteroids}$ to increase with increases in disk timescale. 

\subsection{Asteroid belt mass and size distribution}
The accretion rate extrapolated from N-body simulations depends on the number of objects in the asteroid belt, which in turn depends on the total mass and assumed distribution of radii for the population of asteroids.  Furthermore, the number of bodies contributing to heavy element pollution at any given time, and therefore the median mass of heavy elements in the WD convection zone, varies directly with the accretion rate. In this work, we assumed a solar system mass asteroid belt, with radii spanning $0.5-500 \rm km$ following a distribution of $dN\propto r^{-3.5} dr$.

Due to the continuous nature of asteroid accretion, pollution remains at a relatively stable minimum for the duration of accretion ($\sim 10^{16}$ g total in the DA, $\sim 10^{20}$ g for the DB), with short spikes to greater, observable masses when a particularly massive asteroid accretes. Whether asteroid accretion is observable for long time periods is therefore an `all or nothing' issue, and is very sensitive to how a typical amount of accumulated heavy elements compares to the  detection limit. If the limit is just above the typical mass that can build up from multiple accretions, we will observe only the peaks of the most massive asteroid accretions. However,  if the detectability limit is just below the typical mass of accumulated metals, accretion is observable for the entire time period. 

Given that many polluted white dwarf progenitors are estimated to be more massive than the Sun \citep{Coutu2019}, it is reasonable to assume that many of these systems may have had debris belts more massive than the asteroid belt. If the debris in these more massive belts follows the same collisionally-produced power-law distribution as described for the asteroid belt, there would be correspondingly more objects of any given mass. Assuming the accretion rates per number of available bodies is unchanged, a larger debris belt would increase the total number of accretion events, and therefore increase the typical pollution mass at every point in time, resulting in a larger $T_{\rm asteroids}$. If the moon accretions remain unchanged, this increase in $T_{\rm asteroids}$ would decrease the fraction of pollution that would be due to moons.

\subsection{The impact of planet spacing on moon liberation}
Because moon liberations depend on close encounters between planets, and planet separations determine how quickly systems can become unpacked during stellar evolution, we expect the frequency of moon liberations to vary with planetary system architectures. The three-planet system used throughout \textsection \ref{Section:Asteroids} has spacings of 5-7 mutual Hill radii, with the innermost planet situated at 10 AU. This arrangement of planets is somewhat more tightly packed than most observed systems. Separations for systems detected by the {\it Kepler} satellite generally peak around 14-20 mutual Hill radii \citep{Pu2015, Weiss2018}, however these observed planets are all on orbits interior to $\sim2$ AU, and it is unclear if this trend would directly apply to outer planets. 

One constraint on outer planet spacings is HR 8799 \citep{Marois2008, Marois2010}, which hosts four giant planets ($>5 M_{\rm Jup}$) exterior to 10 AU, and is a likely candidate for a future polluted white dwarf system \citep{Veras2021}. Considering the most stable configuration for these planets \citep{Gozdziewski2020}, separations are approximately 2-3 mutual $R_{\rm H}$. From the sample of directly-imaged exoplanets, \cite{Nielsen2019} found that approximately 9$\%$ of stars more massive than $1.5 M_{\odot}$ could host such massive planets outside of 10 AU.

\begin{figure}
    \centering
    \includegraphics[width=0.5\textwidth]{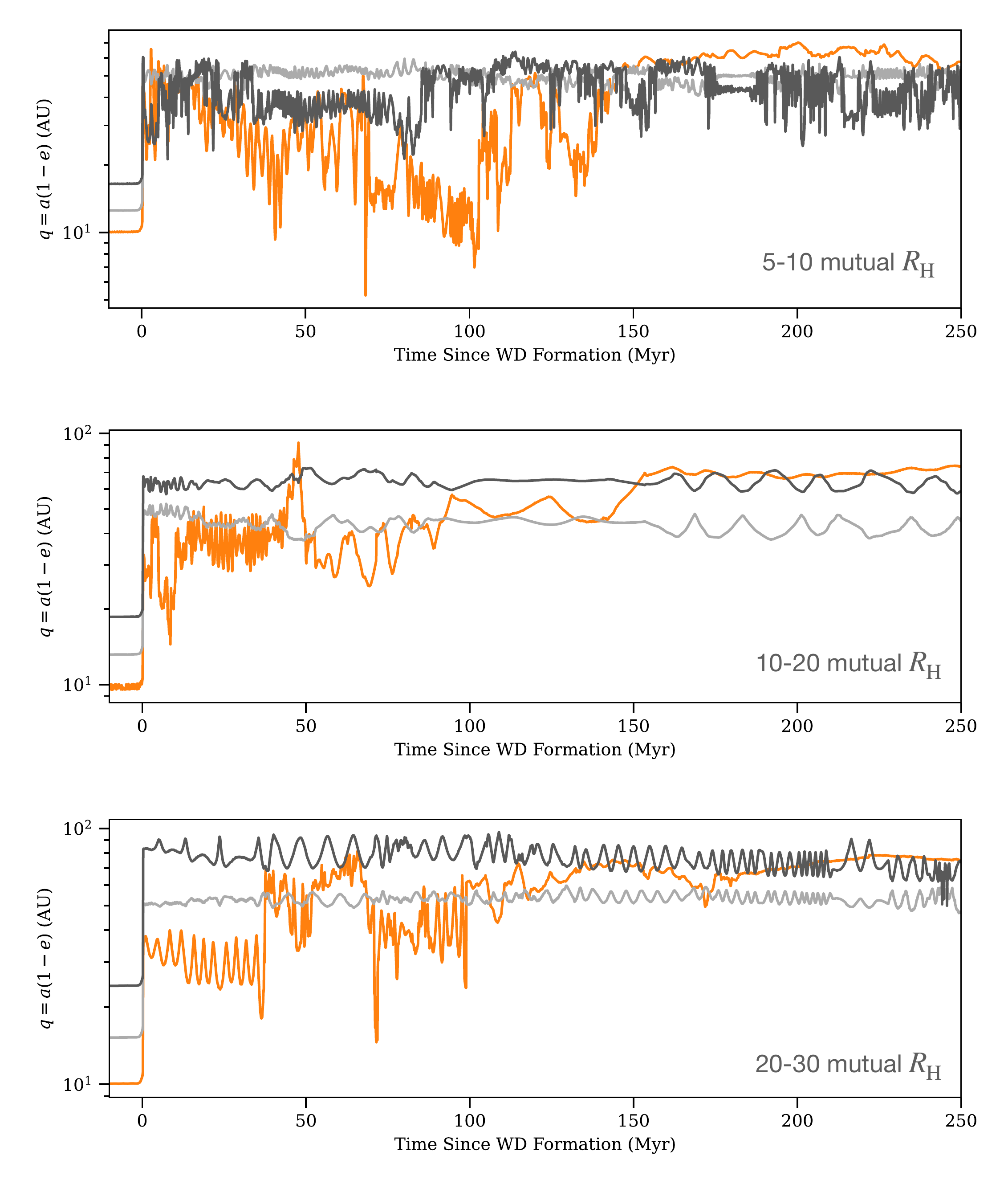}
   \caption{The periapse evolution of three-planet systems for three different ranges of planet spacings in terms of mutual Hill radii: 5-10 (top), 10-20 (middle), and 20-30 (bottom). In each system, the innermost planet is at 10 AU and the spacings between each consecutive planet are randomly chosen from the stated ranges. Planet masses for all three simulations are 1.3, 30.6, and 7.8 $M_{\oplus}$. While the most closely packed system experiences the most orbital crossings between planets, crossings still occur in the more widely spaced systems. }\label{Figure:vary_spacing}
\end{figure}

In Figure \ref{Figure:vary_spacing} we show the orbital crossings that result in simulations with initial planet spacings of 5-10, 10-20 and 20-30 mutual Hill radii. Planet masses for all three simulations are  1.3, 30.6, and 7.8 $M_{\oplus}$, the same as used in \textsection \ref{Section:Asteroids}. These simulations result in 502, 148, and 173 crossings for the 5-10, 10-20, and 20-30 cases, respectively. We find that while the number of crossings varies with planet spacings, crossings do still occur even at spacings more consistent with the majority of observed systems. From this simple comparison we do not anticipate that orbital crossings, and consequently moon liberations, would be entirely eliminated for more widely separated systems. Further study is necessary for more detailed connections between liberations and accretions and planetary architectures.

\subsection{Populations of exomoons}
In this paper, we assumed that moon populations would be similar to those in the solar system. It is possible that moon populations in exoplanet systems do not resemble the solar system. However, as there are no confirmed detections of rocky or icy exomoons, it is difficult to determine how many such moons a typical exoplanet system might have. Additionally, if exomoons in general are not found in all planetary systems, the probabilities of moon accretions derived in this work should be multiplied by the fraction of polluted white dwarf systems that do host exomoons. 
 
A first-order limit on the number of moons in a given WD system should be the placement of moon-hosting planets. \cite{Dobos2021} found that whether exomoons can survive in a stable orbit around a given planet depends on the proximity of the planet to the host star. They conclude that planets on very short period orbits ($< 100$ days) are most limited in the fraction of moons they can retain, while planets with longer periods could retain at least $60\%$ of their original moons. For the purposes of determining the populations of moons available to accrete onto a WD, the longer-period planets are likely more relevant, as the inner planets risk being engulfed by the star in its red giant phase. 

In our three-planet system, planets were originally located 10-13 AU around a 3$M_\odot$ star, with periods of $18-27$ years. According to the Dobos et al. study, these planets could retain about $70\%$ of their moons. The giant planets of the solar system have orbital periods of about 11 to 160 years, and have a similar moon retention fraction of $60-80 \%$. This suggests that whether or not the distributions in mass and semi-major axis of moons in our theoretical three-planet system match those for our solar system's moons, systems resembling our three-planet system should have a large portion of their moons intact and bound to planets when the white dwarf forms. Further constraints on exomoon populations could be made as Transit Timing Variation searches for exomoons progress \citep[e.g.,][]{Teachey2021,Kipping2021}.

\section{Conclusions}
Motivated by the large parent body masses required to explain observed levels of white dwarf pollution, and the recent discovery of beryllium in a white dwarf atmosphere, we have used  N-body simulations and the \cite{Jura2009} accretion model to assess the likelihood that a WD will be polluted by a moon.
We focus this study on the first 200 Myr past WD formation for a planetary system containing three Super-Earth/Neptune-class planets. Extrapolating from asteroid N-body simulations, we find that such a planetary system could sustain an asteroid accretion rate of approximately 1200 objects per Myr. Assuming that the system had a population of moons similar to the regular solar system moons, we find from N-body simulations that we could expect up to about 0.4 moon accretions per 200 Myr. 

Using the population of observed white dwarfs with calcium detections, we find that the pollution must have a calcium mass component of at least $5.8\times 10^{14}$ g to be observable in a DA atmosphere, and $1.4\times10^{18}$ g to be detected in a DB. We use these limits to determine the cumulative fractions of time that moons and asteroids can produce observable levels of pollution. Based on our numerical accretion model, we expect $\sim1\%$ of white dwarf pollution to come from moons as opposed to asteroids. If we consider, as a first-order approximation, that all of the most highly polluted WDs (requiring the most massive parent bodies) are polluted by moons, our parent body mass approach returns a similar statistic, of moons making up about $2\%$ of polluters. 

\acknowledgements
C.M.\ and K.J.W.\ acknowledge support from NSF grants SPG-1826583 and SPG-1823617. EDY acknowledges support from NASA Exoplanets grant  80NSSC20K0270. 


\bibliography{bibli}{}
\bibliographystyle{aasjournal}



\end{document}